\documentclass[final,5p,times,twocolumn]{elsarticle}

\usepackage[table,xcdraw]{xcolor}
\usepackage{url}
\usepackage{enumerate}
\usepackage{amsmath}
\usepackage{graphicx}
\usepackage{array}
\usepackage[inline]{enumitem}
\usepackage[table]{xcolor}
\usepackage{tabularx}
\usepackage{caption}
\usepackage{float}
 \usepackage{tabu}
 \usepackage{multicol}
 \usepackage{multirow}
 \usepackage{longtable}
\usepackage{flushend}
 \usepackage{booktabs}
 \usepackage{adjustbox}
 \usepackage{anyfontsize}
 \usepackage{footnote}
 \makesavenoteenv{tabular}
 \makesavenoteenv{table}
 \usepackage{lipsum}
 \usepackage{listings}
 \usepackage{algorithm2e}
 \usepackage{subfigure}
 \usepackage{booktabs} 
\usepackage{multirow}
\usepackage{listings}

\usepackage{adjustbox}
\usepackage{booktabs}

\usepackage{hyperref}
\makeatletter
\def\study#1{\begingroup
#1%
\def\@currentlabel{\noexpand[#1\noexpand]}%
\phantomsection\label{#1}\endgroup
}
\makeatother

\definecolor{tableShade}{gray}{0.95}
\definecolor{Gray}{gray}{0.7}
\usepackage{spverbatim}

\usepackage{amssymb}

\usepackage{lineno}

\journal{Journal Name}

\begin{document}

\begin{frontmatter}




\title{Completeness and Consistency Analysis for Evolving Knowledge Bases}


\author[polito,ismb]{Mohammad Rifat Ahmmad Rashid}
\author[ismb]{Giuseppe Rizzo}
\author[polito]{Marco Torchiano}
\author[upm]{Nandana Mihindukulasooriya}
\author[upm]{Oscar Corcho}
\author[upm]{Ra\'ul Garc\'ia-Castro}

\address[polito]{Politecino di Torino, Turin, Italy.}
\address[ismb]{Istituto Superiore Mario Boella, Turin, Italy.}
\address[upm]{Ontology Engineering Group, Universidad Polit\'ecnica de Madrid, Boadilla del Monte, Spain}

\begin{abstract}

Assessing the quality of an evolving knowledge base is a challenging task as it often requires to identify correct quality assessment procedures. 
Since data is often derived from autonomous, and increasingly large data sources, 
it is impractical to manually curate the data, and challenging to continuously and automatically assess their quality.
In this paper, we explore two main areas of quality assessment related to evolving knowledge bases: (i) identification of completeness issues using knowledge base evolution analysis, and (ii) identification of consistency issues based on integrity constraints, such as minimum and maximum cardinality, and range constraints.
For completeness analysis, we use data profiling information from consecutive knowledge base releases to estimate completeness measures that allow predicting quality issues. Then, we perform consistency checks to validate the results of the completeness analysis using integrity constraints and learning models.
The approach has been tested both quantitatively and qualitatively by using a subset of datasets from both DBpedia and 3cixty knowledge bases. The performance of the approach is evaluated using precision, recall, and F1 score. From completeness analysis, we observe a 94\% precision for the English DBpedia KB and 95\% precision for the 3cixty Nice KB. We also assessed the performance of our consistency analysis by using five learning models over three sub-tasks, namely minimum cardinality, maximum cardinality, and range constraint. We observed that the best performing model in our experimental setup is the Random Forest, reaching an F1 score greater than 90\% for minimum and maximum cardinality and 84\% for range constraints. 

\end{abstract}


\end{frontmatter}

\section{Introduction}\label{sec:intro}

In recent years, numerous efforts have been put towards sharing Knowledge Bases (KBs) in the Linked Open Data (LOD) cloud\footnote{\url{http://lod-cloud.net}}. This has led to the creation of large corpora, making billions of RDF\footnote{\url{https://www.w3.org/RDF}} triples available from different domains such as Geography, Government, Life Sciences, Media, Publication, Social Networking, and User generated data.
These KBs evolve over time: their data instances and schemas are updated, extended, revised and refactored \cite{gottron2014perplexity}. Unlike in more controlled types of knowledge bases, the evolution of KBs exposed in the LOD cloud is usually unrestrained~\cite{debattistaevaluating}, which may cause data to suffer from a variety of quality issues, at both schema level and data instance level. Considering the aggregated measure of conformance, the empirical study carried out by Debattista \textit{et al.}~\cite{debattistaevaluating} shows that datasets published in the LOD cloud have reasonable overall quality, but significant issues remain concerning different quality metrics, such as data provenance and licensing.
Therefore, by looking at individual metrics, we can explore certain aspects, for example data quality issues in the data collection or integration processes. 

Data quality relates to the perception of the ``fitness for use'' in a given context~\cite{tayi1998}.
One of the common tasks for data quality assessment is to perform a detailed data analysis with data profiling~\cite{Olson2003}. Data profiling is usually defined as the process of examining data to collect statistics and provide relevant metadata about the data~\cite{Naumann2014}.
Based on the information we gather from data profiling, we can thoroughly examine and understand a KB, its structure, and its properties before using the KB. Various approaches have been developed for KB quality assessment based on manual, semi-automatic, and automated approaches. For example, Flemming's~\cite{flemming2010quality} data quality assessment approach evaluate data quality scores based on manual user input for data sources. RDFUnit\footnote{\url{http://github.com/AKSW/RDFUnit}} is a tool centered around the definition of integrity constraints for automatic validation tasks. These approaches can ensure an appropriate quality assessment procedure, but it is challenging to continuously and automatically access a evolving KB~\cite{papavasileiou2013high}. 
Various approaches are based on low-level rules and programs, which require a significant user involvement. 
Furthermore, in the current literature less focus has been given into studying the evolution of a knowledge base to detect quality issues.

Ellefi \textit{et al.} \cite{ellefi2017rdf} explored data profiling features of KB evolution by considering the use cases presented by K{\"a}fer \textit{et al.}~\cite{kafer2013observing}.
KB evolution analysis using data profiling features can help to understand the changes applied to an entire KB or parts of it. It has multiple dimensions regarding the dataset update behavior, such as frequency of change, change patterns, change impacts, and causes of change. More specifically, by exploring KB evolution, we can capture those changes that happen often; or changes that the curator wants to highlight because they are useful or interesting for a specific domain or application; or changes that indicate an abnormal situation or type of evolution~\cite{nishioka2016information,papavasileiou2013high,rifat2018}.

The KB evolution can directly impact the data integration tasks (e.g., matching, linking), that may lead to incomplete or incorrect results~\cite{pernellerdf}. For example, Wikipedia has grown into one of the central hubs of knowledge sources, and it is maintained by thousands of contributors. It evolves each day with contributions from editors all over the world.
DBpedia is a crowd-sourced knowledge base and extracts structured information from various Wikipedia projects. 
This extracted data might have quality problems because either they are mapped incorrectly or the source information itself is incorrect~\cite{mihindukulasooriya2016collaborative}.

Considering the level of changes and complexity, KB evolution can be explored based on both simple changes at low-level and complex changes at high-level~\cite{papavasileiou2013high}.
Low-level changes are easy to define and have several interesting properties~\cite{papavasileiou2013high}. For example, low-level change detection in its simplest form performs two operations, detection of addition and deletion, which determine individual resources that were added or deleted in a KB \cite{papavasileiou2013high,volkel2006semversion}. 
However, a detailed low-level and automated analysis is computationally expensive and might result into a huge number of fine-grained issue notifications~\cite{debattista2016luzzu}. Such amount of information might cause an information overload for the receiver of the notifications. On the contrary, high-level analysis captures the changes that indicate an abnormal situation and generates results that are intuitive enough for a human user. However, high-level analysis requires fixed set of requirements (\textit{i.e.}, integrity constraints) to understand underlying changes happened in the dataset~\cite{papavasileiou2013high}. A data quality assessment approach using high-level change detection may lead to increasing the number of false positive results if the version of a KB is deployed with design issues, such as erroneous schema definitions~\cite{rifat2018}.

In particular, a knowledge base is defined to be consistent if it does not contain conflicting or contradictory data~\cite{hogan2010weaving}. Without proper data management, the dataset in an evolving KB may contain consistency issues~\cite{meimaris2015framework}. When a schema is available with integrity constraints, the data usually goes through a validation process that verifies the compliance against those constraints. Those integrity constraints encapsulate the consistency requirements of data in order to fit for a set of use cases.
Considering the limitations of high-level change detection and the changes present at the schema level, integrity constraints based consistency analysis can help to validate the high-level analysis result.
Traditionally, in databases, constraints are limitations incorporated in the data that are supposed to be satisfied all the time by instances~\cite{abiteboul1995foundations}. They are useful for users to understand data as they represent characteristics that data naturally exhibits~\cite{liddle1993cardinality}. In practical settings, constraints are used for three main tasks: \textit{(i)} specifying properties that data should hold; \textit{(ii)} handle contradictions within the data or with respect to the domain under consideration; or \textit{(iii)} for query optimization.
Taking into account ontologies for validation tasks, there are, however, significant theoretical and practical problems.
For example, the OWL W3C Recommendation, based on Description Logic and the Open World Assumption, was designed for inferring new knowledge rather than for validating data using axioms. Reasoners and validators have different functions, \textit{i.e.}, a reasoner is used for inferring new knowledge, even though it may find some inconsistencies as well, while a validator is used for finding violations against a set of constraints. It is a tedious, time-consuming, and error-prone task to generate such validation rules manually. Some of the validation rules can be encoded into the ontology, but it still requires a lot of manual effort. This leads to the need for an approach for inducing such validation rules automatically. Such rules can be represented in the form of RDF shapes by profiling the data and using inductive approaches to extract the rules. Other use cases for inducting shapes include describing the data (which is helpful in validating the completeness analysis results).

In this work, based on the high-level change detection, we aim to analyze completeness issues in any knowledge base. In particular, we address the challenges of completeness analysis for evolving KB using data profiling features. We explore completeness of KB resources using metrics that are computed using KB evolution analysis. The first hypothesis (H1) that has guided our investigation is:

\emph{Data profiling features can help to identify completeness issues.}

We formulate this research goal into the following research question:

\textbf{RQ1:} \emph{To what extent the periodic profiling of an evolving KB can contribute to unveil completeness issues?} 

In response to RQ1, we explore the completeness analysis approaches similar to the work presented in \cite{rifat2018}. In particular, we explore multiple data profiling features at the class level and at the property level to define completeness quality measures. For the measurement functions, we use basic summary statistics (i.e. counts and diffs) over entities from periodic KB releases.

To validate the completeness analysis results, we present an experimental analysis that is based on a qualitative and constraints-based validation approach. We propose constraints based feature extraction approach to address the challenges of consistency issues identification in an evolving KB. For constraints-based consistency evaluation, we derived the second hypothesis (H2):

\emph{Learning models can be used to predict correct integrity constraints using the outputs of the data profiling as features.}

We present this research goal into the following research question:

\textbf{RQ2:} \emph{How can we perform consistency checks using integrity constraints as predictive features of learning models?}

To address RQ2, we use KB data profiling information to generate integrity constraints in the form of SHACL~\cite{shacl} RDF shapes. More specifically, we learn what are the integrity constraints that can be applicable to a large KB by instructing a process of statistical analysis for feature extraction that is followed by a learning model. Furthermore, we performed qualitative analysis to validate the proposed hypothesis by manually examining the results of the completeness analysis. 

The remainder of this paper is organized as follows:

\begin{itemize}	
    \item In Section~\ref{sec:Background}, we present background and motivational examples that demonstrate important key elements of our quality assessment and validation approach;
    
    \item In Section~\ref{sec:related}, we present the related work focusing on linked data dynamics, knowledge base quality assessment, and knowledge base validation;
    
    \item In Section~\ref{sec:metrics}, we explore the concept of KB evolution analysis to drive completeness measurement functions and the process of integrity constraints based on shape induction for consistency analysis; 
    
    \item In Section~\ref{sec:approach}, we present our data driven completeness and consistency analysis approach;

   \item In Section~\ref{sec:experimentalAnalysis}, we present an experimental analysis based on two KBs, namely DBpedia and 3cixty Nice. Furthermore, we considered both English and Spanish versions of DBpedia KB;
   
   \item In Section~\ref{sec:discussion}, we discuss the hypothesis, the research questions and insights gathered from the experimentation. We also list potential threats emerged while testing the proposed approach;
   
   \item In Section~\ref{sec:conclusion}, we conclude by revisiting each research question and outlining future research endeavours.
\end{itemize}

\section{Background and Motivation}\label{sec:Background}

In this work, we explored two KBs namely, the 3cixty Nice~\cite{3cixty} and DBpedia~\cite{Dbpedia}. Here we report a few common prefixes used over the paper:

\begin{itemize}
	\item DBpedia ontology URL\footnote{\url{http://dbpedia.org/ontology/}} prefix: \textit{dbo};
	\item DBpedia resource URL\footnote{\url{http://dbpedia.org/resource/}} prefix: \textit{dbr};
	\item FOAF Vocabulary Specification URL\footnote{\url{http://xmlns.com/foaf/0.1/}} prefix: \textit{foaf};
	\item Wikipedia URL\footnote{\url{https://en.wikipedia.org/wiki/}} prefix: \textit{wikipedia-en};
	\item 3cixty Nice event type URL\footnote{\url{http://linkedevents.org/ontology}} prefix: \textit{lode};
	\item 3cixty Nice place type URL\footnote{\url{http://www.ontologydesignpatterns.org/ont/dul/DUL.owl}} prefix: \textit{dul}.
\end{itemize}

In this section, we present an overview of the two main research areas: \textit{(i)} identification of completeness issues using KB evolution analysis, and \textit{(ii)} identification of consistency issues based on integrity constraints. Also, we outline the approaches for gold standard creation and learning models.

\subsection{Identification of completeness issues using KB evolution analysis}

For a specific context of use, a completeness issue is associated with an entity having all expected attributes~\cite{SQUARE}. More specifically, it is associated with the quality issues related to missing entities or missing properties of a knowledge base. This may happen because of an unexpected deletion or because of data source extraction errors.

Fine-grained completeness analysis based on low-level changes brings substantial data processing challenges~\cite{papavasileiou2013high,pernellerdf}. More specifically, low-level change detection compares the current dataset version with the previous one and returns the delta containing the added or deleted entities. For example, two DBpedia versions -- 201510 and 201604 -- have the property \textit{dbo:areaTotal} in the domain of \textit{dbo:Place}. Low-level changes can help to detect added or deleted instances for \textit{dbo:Place} entity type. One of the main requirements for quality assessment would be to identify the completeness of \textit{dbo:Place} entity type with each KB releases. Low-level changes can help only to detect missing entities with each KB release. Such as those entities missing in the 201604 version (e.g. dbr:A\_R\'{u}a,  dbr:Sandi\'{a}s, dbr:Coles\_Qurense). Furthermore, these instances are automatically extracted from Wikipedia Infobox keys. 
We track the Wikipedia page from which DBpedia statements were extracted.
These instances are present in the Wikipedia Infobox as Keys but missing in the DBpedia 201604 release. It is not feasible to manually check all such missing entities or attributes. Thus, because of the large volume of the dataset, it is a tedious, time-consuming, and error-prone task to perform such quality assessment manually.

The representation of changes at low-level leads to syntactic and semantic deltas~\cite{volkel2006semversion} from which it is more difficult to get insights to complex changes or changes intended by a human user. On the contrary, high-level changes can more efficiently capture the changes that indicate an abnormal situation and generates results that are intuitive enough for a human user. High-level changes from the data can be detected using statistical profiling. For example, total entity count of \textit{dbo:Place} type for two DBpedia versions -- 201510 and 201604 -- is 1,122,785 and 925,383 where the entity count of 201604 is lower than 201510. This could indicate an imbalance in the data extraction process without fine-grained analysis. 

\begin{figure}[!ht]
  \centering
  \includegraphics[width=.5\linewidth]{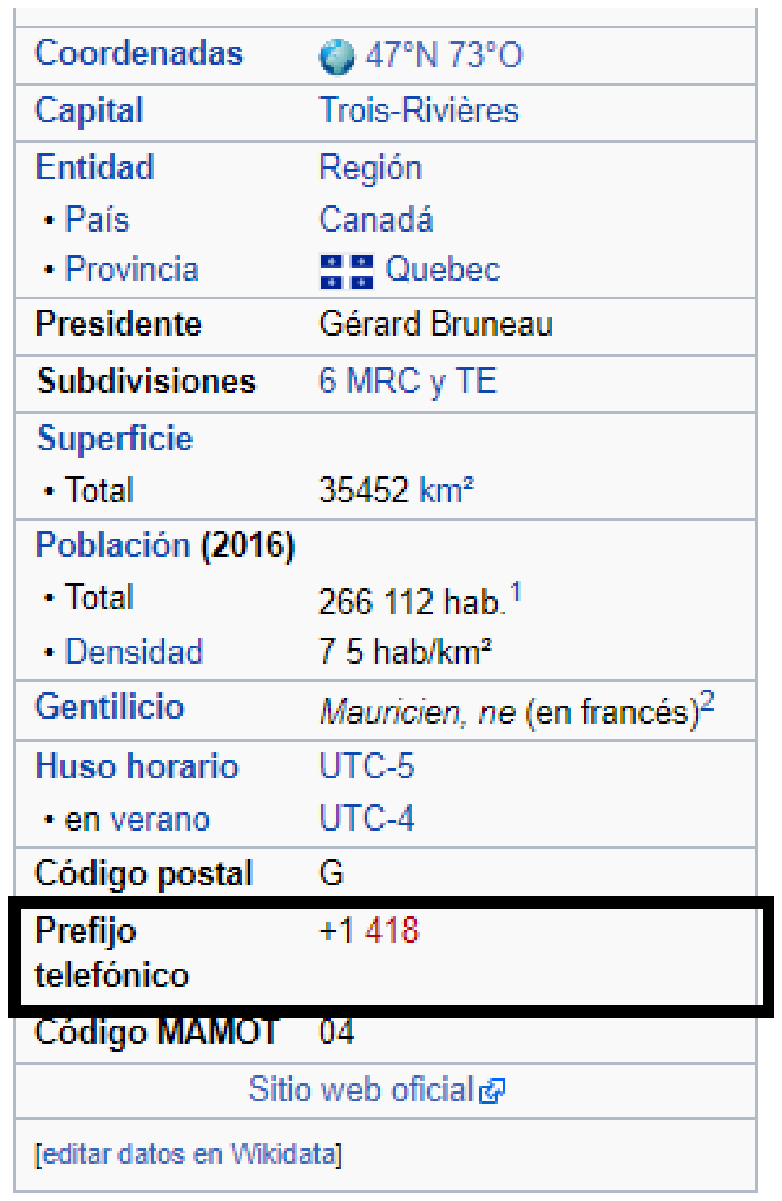}
  \caption{Example of incomplete Wikipedia data.}
  \label{compExamp}
\end{figure}

As an example, let us consider a DBpedia ES\footnote{\url{http://es.dbpedia.org}} entity \textit{dbo:Place/prefijoTelef\'{o}nicoNombre}:\textit{Mauricie}.\footnote{\url{http://es.dbpedia.org/page/Mauricie}}
When looking at the source Wikipedia page,\footnote{\url{https://es.wikipedia.org/wiki/Mauricie}} we observe that, as shown in Figure~\ref{compExamp}, the infobox reports a ``Prefijo telef\'{o}nico'' datum. The DBpedia ontology includes a \textit{dbo:Place/prefijoTelef\'{o}nicoNombre}, and several other places have that property, but the entity we consider is missing that information. 

While it is generally difficult to spot that kind of incompleteness, for the case under consideration it is easier because that property was present for the entity under consideration in the previous version of DBpedia ES~\cite{Dbpedia} i.e. the 2016-04 release. It is a completeness issue introduced by the evolution of the knowledge base. It can be spotted by looking at the frequency of predicates inside for an entity type. In particular, in the release of 201610 there are $55,387$ occurrences of the \textit{dbo:Place/prefijoTelef\'{o}nicoNombre} predicate over $356,479$ \textit{dbo:Place} entity type, while in the previous version (201604) they were $56,109$ out of $657,481$ \textit{dbo:Place} entities.


Based on the linked data dynamicity behaviour \cite{meimaris2015framework,ellefi2017rdf,kafer2013observing}, we can assume that the growth of the entities in a mature KB ought to be stable. In this aspect, another completeness issue relates to entities that were present in the previous knowledge base releases, but disappeared from more recent ones.
As an example, let us consider a 3cixty Nice entity of type \textit{lode:Event} that has as label: ``Mod\'eliser, piloter et valoriser les actifs des collectivit\'es et d'un territoire gr\^ace aux maquettes num\'eriques: retours d'exp\'eriences et bonnes pratiques''.
This entity happened to be part of the 3cixty Nice KB since it has been created the first time, but in a subsequent release it got removed even though it should not. Such a problem is generally complex to be traced manually because it requires a per-resource check over the different releases. It can, instead, be spotted by looking at the total frequency of entities of a given resource type. 

\begin{figure}[!ht]
  \includegraphics[width=\linewidth]{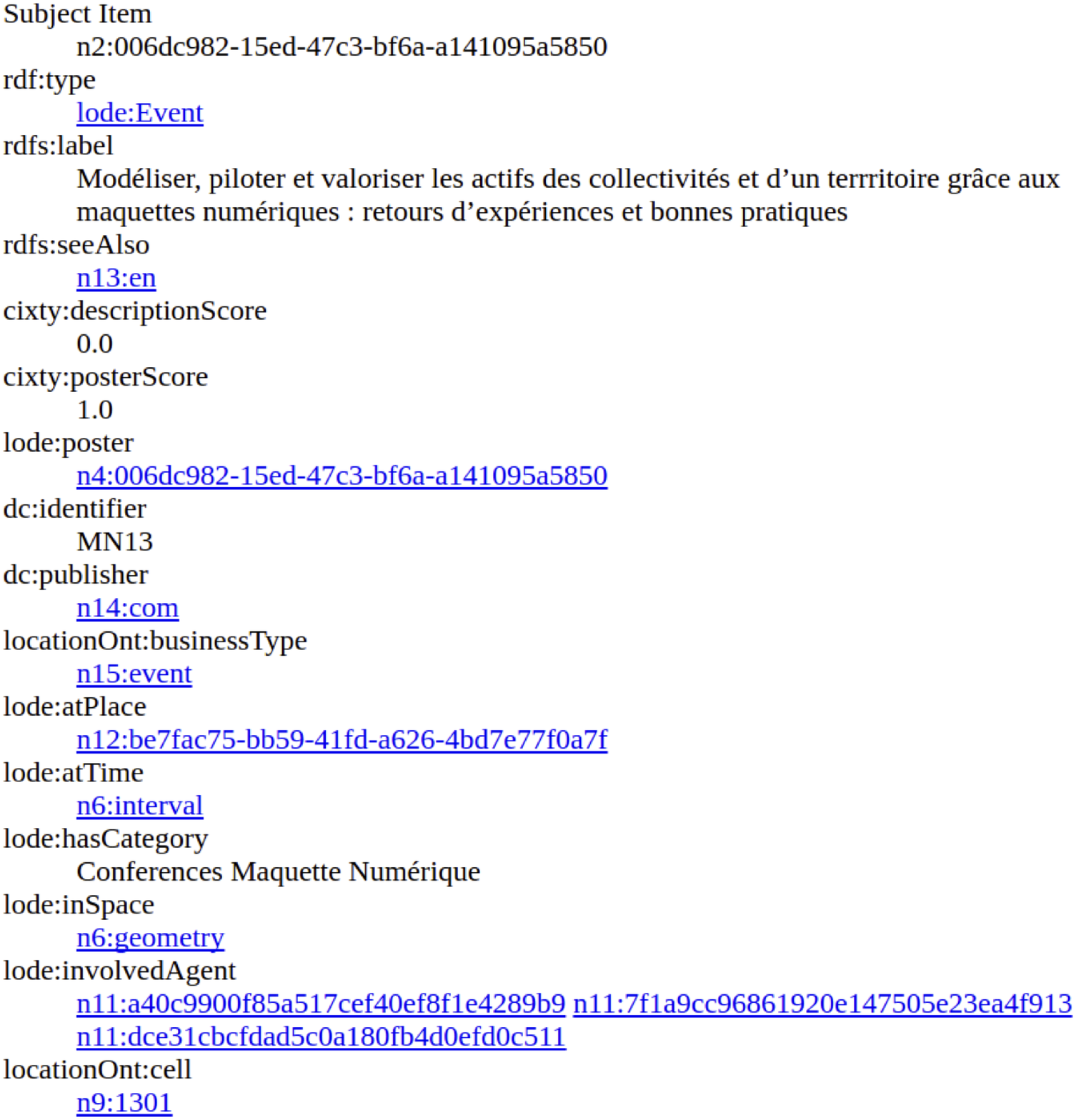}
  \caption{Example of a 3cixty Nice KB entity that unexpectedly disappeared from the release of 2016-06-15 to the other 2016-09-09.}
  \label{persistencyExamp}
\end{figure}

\subsection{Identification of consistency issues based on integrity constraints}

Another issue of unrestrained KB evolution is the unavailability of explicit schema information that precisely defines the types of entities and their properties~\cite{nishioka2016information}.
In a KB, when an ontology is available with TBox axioms, which define the conceptualization of the domain, a reasoner can be used to verify whether the dataset is consistent with the domain by verifying the axioms defined in the ontology~\cite{paulheim2016fast}.
The empirical study presented by Mihindukulasooriya \textit{et al.}~\cite{mihindukulasooriya2016collaborative} 
pointed out that changes in the ontology depend on the development process and the community involved in the creation of the knowledge base. They also pointed out the drawbacks of finding practical guidelines and best practices for ontology based evaluation. 
Taking into account availability of schema with integrity constraints, the data usually goes through a validation process that verifies the compliance against those constraints. Those integrity constraints encapsulate the consistency requirements of data in order to fit for a set of use cases. For example, in a relational database, the integrity constraints are expressed in a data definition language (DDL), and the database management system (DBMS) ensures that any data inserted into the entire database will not lead to any inconsistency.

The validation of entities in a KB is not done in the same manner as in traditional database management systems due to the lack of a language for expressing constraints or having less restrictive generic models suitable for wider use and not for specific use cases. Furthermore, most of the ontologies do not have rich axioms that could help to detect inconsistencies in data~\cite{paulheim2016fast}. Further, most of the schema information about RDF data is only available in the form of OWL ontologies that are most suited for entailment rather than validation. In this account, we can assume that practical use cases that utilize RDF data need the validation of integrity constraints.
Larger knowledge bases, such as DBpedia, lack the precise definition of integrity constraints, and it is a tedious task to create these constraint definitions from scratch manually. In DBpedia KB~\cite{auer2007dbpedia} (version 2016-04), a person should have exactly one value for the "dbo:birthDate" property or the values of the "dbo:height" property should always be a positive number. The instances of the Person class have more than 13,000 associated properties (including dbo, DBpedia ontology properties and dbp, auto-generated properties from Wikipedia infobox keys).
Taking into account ontologies for consistency analysis, they are usually designed for entailment purposes rather than for assessment and their representation often lacks the granular information needed for validating constraints in the data~\cite{gayo2017validating}. This leads to the need of automatic consistency analysis for evolving KBs.

\subsection{Gold Standard Creation}\label{sec:goldStandard}

KBs may contain errors, thus the profiling results cannot be considered as a gold standard. There are different strategies to evaluate a dataset. In the following, we present three common strategies when dealing with a knowledge base~\cite{paulheim2017knowledge}:

\textit{(i)}\textit{ Silver Standard:}
this strategy is based on the assumption that the given KB is already of reasonable quality. The silver standard method is usually applied to measure the performance of knowledge graph by analyzing how well relations in a knowledge graph can be replicated. Although this strategy is suitable for large-scale data, it can produce less reliable results~\cite{groza2013using, kang2012training}.

\textit{(ii) Gold standard:} this strategy is based on turning the observations in a set of gold data points by human annotators. 
In this context, gold standard is indeed suitable for our approach since we can obtain gold insights of the completeness measurement results, however it is very expensive if the annotation load is large.

\textit{(iii) Partial gold standard:} in this strategy, a small subset of external graphs, entities or relations are selected as validation criteria and they, then, are manually labeled~\cite{groza2013using}. This helps reducing the number of candidates that an annotator will process.

\subsection{Learning Models}\label{sec:ML}

We considered consistency analysis of instances using RDF shape induction as a classification problem. Typically a classification learning model maps observations (samples) to a set of possible categories (classes)~\cite{witten2016data}. 
For example, the minimum cardinality value of an entity type is an observation for its relevant attributes (features). For selecting a suitable learning model for our problem, we investigated the following research question: \textit{"Which learning model is the most adequate for consistency analysis using data profiling information as predictive features?"}. 
In order to answer this question, we evaluate the performance of predictive features using five classical learning models. These learning models are chosen considering five categories of machine learning algorithms~\cite{witten2016data}: \textit{(i)} Neural Networks, \textit{(ii)} Bayesian, \textit{(iii)} Instance Based, \textit{(iv)} Support Vector Machine, and \textit{(v)} Ensemble. Following we present details of those tested in this work.

\textit{Multilayer Perceptron}~\cite{hornik1989multilayer}: a feed forward Neural Network consisting of at least three layers of neurons with a non-linear activation function: one for inputs, one for outputs and one or more hidden layers. Training is carried out through back propagation.

\textit{Naive Bayes}~\cite{domingos1997optimality}: is a simple probabilistic classifier. The core concept is based on the Bayes theorem~\cite{domingos1997optimality}. Generally, naive bayes classifiers are based on the assumption that features are independent with each other.

\textit{k-Nearest Neighbors (k-NN)}~\cite{aha1991instance}: is an instance-based learning algorithm. It locates the \textit{k}-nearest instances to the input instance and determines its class by identifying the single most frequent class label. It is generally considered not tolerant to noise and missing values. Nevertheless, it is highly accurate, insensitive to outliers and works well with both nominal and numerical features.

\textit{Support Vector Machines (SVM)}~\cite{cortes1995support}: it conceptually implements the following idea: input vectors are non-linearly mapped to a very high dimensional feature space. In this feature space a linear decision surface is constructed. Special properties of the decision surface ensures high generalization ability of the classifier.

\textit{Random Forest}~\cite{pfahringer2010random}: it creates many classification trees. To classify a new object from an input vector, it maps the input vector down each of the trees in the forest. Each tree gives a classification, and we say the tree ``votes'' for that class. The forest chooses the classification having the most votes (over all the trees in the forest).


In our modeling phase, we applied a \textit{k-fold cross validation} \cite{witten2016data} 
to reduce the variance of a performance score. In the \textit{k-fold cross validation} setup, \textit{k} is the number of splits to make in the dataset. We choose value of \textit{k}=10. This results in splitting the dataset into 10 portions (10 folds) and runs the learning model 10 times. For each algorithm, the training runs on 90\% of the data and testing on the left 10\%. 
With \textit{k} value of 10, it uses each data instance as a training instance exactly 9 times and test instance 1 time.

We also adopted general classification performance evaluation measures such as precision, recall, and F1 score~\cite{sokolova2009systematic}. Evaluation of the classification performance is based on considering one of the output classes as the positive class and defining: \textit{(i)} true positives (TP): the number of samples correctly labeled as in the positive class; \textit{(ii)} false positives (FP): the number of samples incorrectly labeled as in the positive class; \textit{(iii)} true negatives (TN): the number of samples correctly labeled as not in the positive class; \textit{(iv)} false negatives (FN): the number of samples incorrectly labeled as not in the positive class.

We present the formulas of the aforementioned metrics:

\begin{description}
\item[Precision ($P$):] it is based on positive predictive value and it defined as $P=\frac{TP}{TP+FP}$; 
\item[Recall ($R$):] it is related to true positive rate also know as sensitivity and it is defined as $R=\frac{TP}{TP+FN}$; 
 
\item[F1 Score ($F1$):] it is a measure of test accuracy and it is defined as the harmonic mean of precision and recall: $F1=\frac{2*P*R}{P+R}$. 
 
\end{description}

\section{Related Work}
\label{sec:related}

This section provides an overview of the state-of-the-art in the context of knowledge base quality assessment approaches. The research activities related to our approach fall into three main areas: \textit{(i)} Linked Data Dynamics, \textit{(ii)} Knowledge Base Quality Assessment, and \textit{(iii)} Knowledge Base Validation.

\subsection{Linked Data Dynamics}

Taking into account changes over time, every dataset can be dynamic. Considering linked data dynamics, a comparative analysis is present by Umbrich \textit{et al.}~\cite{umbrich2010towards}. The authors analyzed entity dynamics using a labeled directed graph based on LOD, where a node is an entity that is represented by a subject. In addition, Umbrich \textit{et al.} \cite{umbrich2010dataset} presented a comprehensive survey based on technical solutions for dealing with changes in datasets of the Web of Data.
Furthermore, K{\"a}fer \textit{et al.}~\cite{kafer2013observing} designed a Linked Data Observatory to monitor linked data dynamics. The authors setup a long-term experiment to monitor the two-hop neighborhood of a core set of eighty thousand diverse Linked Data documents on a weekly basis. Furthermore, linked data dynamics is considered using five use cases: synchronization, smart caching, hybrid architectures, external-link maintenance, and vocabulary evolution and versioning.

The work presented by Papavasileiou \textit{et al.}~\cite{papavasileiou2013high} explored high-level change detection in RDF(S) KBs by addressing change management for RDF(S). The authors explored the data management issues in KBs where data is maintained by large communities, such as scientists or librarians, who act as curators to ensure high quality of data.
Such curated KBs are constantly evolving for various reasons, such as the inclusion of new experimental evidence or observations, or the correction of erroneous conceptualizations. Managing such changes poses several research problems, including the problem of detecting the changes (delta) among versions of the same KB developed and maintained by different groups of curators, a crucial task for assisting them in understanding the involved changes. The authors addressed this problem by proposing a language for change detection that allows the formulation of concise and intuitive deltas. Similarly, in our work, we explore the deltas present in consecutive KB releases using data profiling.

In \cite{mihindukulasooriya2016collaborative}, Mihindukulasooriya \textit{et al.} presented an empirical analysis of the ontologies that were developed collaboratively to understand community-driven ontology evolution in practice.
The authors have analyzed, how four well-known ontologies (DBpedia, Schema.org, PROV-O, and FOAF) have evolved through their lifetime and they observed that quality issues were due to the ontology evolution. Also, the authors pointed out the need for having multiple methodologies for managing changes. The authors summarize that the selected ontologies do not follow the theoretical frameworks found in the literature. Further, the most common quality problems caused by ontology changes include the use of abandoned classes and properties in data instances and the presence of duplicate classes and properties. Nevertheless, this work is not focused on KB evolution analysis for completeness analysis but rather on how changes in the ontology affect the data described using those ontologies. Klein \textit{et al.}~\cite{klein2002ontology} studied the ontology versioning in the context of the Web. The authors looked at the characteristics of the release relation between ontologies and at the identification of online ontologies. Then, a web-based system is introduced to help users to manage changes in ontologies. Similarly, Pernelle \textit{et al.}~\cite{pernellerdf} presented an approach that detects and semantically represents data changes in knowledge bases. However, ontologies description for KBs are not always available~\cite{mihindukulasooriya2016collaborative}. In this work, we focus on KB evolution analysis using data profiling at the data instance level to address the issues concerning unavailability of ontology descriptions.. 

In~\cite{nishioka2016information}, Nishioka \textit{et al.} presented a clustering technique over the dynamics of entities to determine common temporal patterns. The quality of the clustering is evaluated using entity features such as the entities, properties, RDF types, and pay-level domain. Besides, the authors investigated to what extent entities that share a feature value change over time.
In this paper, we explore linked dynamic data features for detecting completeness issues. Instead of using a clustering technique~\cite{nishioka2016information} based on the temporal pattern of entities, we focus on exploring the evolution analysis as a classification problem to detect completeness issues.


\subsection{Knowledge Base Quality Assessment}

Knowledge Base quality assessment is a largely investigated research field, and many approaches to data quality management have been proposed. 
There exists a large number of data quality frameworks and tools based on manual, crowd-sourced, and automatic approaches. 
In this section, we review literature that analyze the quality of various aspects of KBs. 

\textit{Comprehensive Studies.} A comprehensive overview of the RDF data profiling is presented by 
Ellefi \textit{et al.}~\cite{ellefi2017rdf}. The authors explored the RDF data profiling feature, methods, tools, and vocabularies. 
Furthermore, the authors presented dataset profiling in a taxonomy and illustrated the links between the dataset profiling and feature extraction approaches.
Ellefi \textit{et al.} organized dataset profiling features into seven top-level categories: 
\begin{enumerate*}
	\item General;
	\item Qualitative;
	\item Provenance;
	\item Links;
	\item Licensing;
	\item Statistical;
	\item Dynamics.
\end{enumerate*}
The authors considered linked data dynamics as profiling features using the study presented by K{\"a}fer \textit{et al.}~\cite{kafer2013observing}. Similarly, in this work, we explore the concepts regarding qualitative, statistical, and dynamic features. Also, based on Ellefi~\textit{et al.}~\cite{ellefi2017rdf} study, we explore the dynamic features to perform completeness analysis. 

Considering the data quality methodologies applied to linked open data (LOD), a comprehensive systematic literature review is presented by Zaveri \textit{et al.}~\cite{zaveri2016quality}. The authors have extracted 26 quality dimensions and a total of 110 objective and subjective quality indicators.  Zaveri \textit{et al.} organized the linked data quality dimensions into the following categories,
\begin{enumerate*}
	\item Contextual dimensions;
	\item Trust dimensions;
	\item Intrinsic dimensions;
	\item Accessibility dimensions;
	\item Representational dimensions;
	\item Dataset dynamicity dimensions.
\end{enumerate*}  
Furthermore, dataset dynamicity dimensions are explored using three quality dimensions:
\begin{enumerate*}
	\item Currency: speed of information update regarding information changes;
	\item Volatility: length of time which the data remains valid;
	\item Timeliness: information is available in time to be useful.
\end{enumerate*} 
The work presented in this paper is related to contextual and dataset dynamicity dimensions. More concretely, the completeness and consistency is associated with the contextual dimensions, and the dataset evolution is related to the dataset dynamicity dimensions.   

\textit{Quality Assessment Frameworks.} Taking into account data quality analysis using manual approaches, Bizer \textit{et al.}~\cite{bizer2009quality} presented Web Information Quality Assessment Framework (WIQA). The WIQA - Information Quality Assessment Framework is a set of software components that empowers information consumers to employ a wide range of different information quality assessment policies to filter information from the Web. 
This framework employs the Named Graphs data model for the representation of information together with quality-related meta-information and uses the WIQA-PL\footnote{\url{http://wifo5-03.informatik.uni-mannheim.de/bizer/wiqa/}} policy language for expressing information filtering policies. WIQA-PL policies are expressed in the form of graph patterns and filter conditions. WIQA can be used to understand the intended changes present in a KB by applying graph patterns and filtering conditions. In this work, instead of a static version of a KB, we plan to explore multiple versions of KB using WIQA policy.

Using provenance metadata information, Mendes \textit{et al.}~\cite{mendes2012sieve} presented Sieve framework that uses user configurable quality specification for quality assessment and fusion method. Sieve is integrated as a component of the Linked Data Integration Framework (LDIF).\footnote{\url{ http://ldif.wbsg.de/}}
In particular, Sieve uses the LDIF provenance metadata and the user configured quality metrics to generate quality assessment scores. A set of Linked Data quality assessment measures are proposed as:
\begin{enumerate*}
	\item Intensional completeness;
	\item Extensional completeness;
	\item Recency and reputation;
	\item Time since data modification;
	\item Property completeness;
	\item Property conciseness;
	\item Property consistency.
\end{enumerate*}
In this work, instead of using provenance metadata and the user configured quality metrics, we explore completeness using dynamic linked data profiling features presented by Ellefi \textit{et al.}~\cite{ellefi2017rdf}.

In~\cite{kontokostas2014test}, Kontokostas \textit{et al.} proposed a methodology for test-driven quality assessment of Linked Data. The authors formalized quality issues and employed SPARQL query templates, which are instantiated into quality test queries. Also, the authors presented RDFUnit\footnote{\url{https://github.com/AKSW/RDFUnit}} a tool centered around schema validation using test-driven quality assessment approach. RDFUnit runs automatically based on a schema and manually generates test cases against an endpoint. RDFUnit has a component that turns RDFS axioms and simple OWL axioms into SPARQL queries that check for data that does not match the axiom.
In contrast, in this study, we aim to learn the constraints (which might not be explicitly stated as RDFS or OWL axioms) as RDF Shapes. Although the overall objectives are similar considering RDF shape induction to this work, for completeness analysis we mainly explore KB evolution analysis. Furthermore, in our approach, we primarily use data profiling information as the input for the process. Results from the consistency analysis can be extended by using RDFUnit for further validation.

In~\cite{debattista2016luzzu}, Debattista \textit{et 
al.} presented a conceptual methodology for assessing Linked Datasets and proposed Luzzu, a framework for Linked Data Quality Assessment. Luzzu is based on four major components:
\begin{enumerate*}
\item An extensible interface for defining new quality metrics; 
\item An interoperable, ontology-driven back-end for representing quality metadata and quality problems that can be re-used within different semantic frameworks; 
\item Scalable dataset processors for data dumps, SPARQL endpoints, and big data infrastructures;  
\item A customisable ranking algorithm taking into account user-defined weights.
\end{enumerate*} 
Luzzu is a stream-oriented quality assessment framework that focuses on data instance-centric measurement of a user-defined collection of quality metrics. 
The validation metrics require users to write Java code for implementing checks. In this work, we perform completeness analysis based on high-level change detection to identify any problem in the data processing pipeline.
Furthermore, various research works explored the importance of quality metrics in a probabilistic and deterministic settings. Debattista \textit{et al.}~\cite{debattista2015quality} explored probabilistic techniques such as Reservoir
Sampling, Bloom Filters and Clustering Coefficient estimation for implementing a broad set of data quality metrics in an approximate but sufficiently accurate way. 
In addition, various research works put emphasis on the problem of error detection in a KB. For example, distance-based outlier detection by Debattista \textit{et al.}~\cite{debattista2016preliminary} and error detection in relation assertions by Melo \textit{et al.}~\cite{melo2017detection} gave more focus towards error detection in schemas. 
The core of the study is similar considering error detection in a KB, the focus of this study is to identify completeness and consistency issues using various data profiling features.

\textit{Crowdsourcing.} A crowd-sourcing quality assessment approach can be used to understand the intended changes by stakeholders due to KB updates. Acosta \textit{et al.}~\cite{acosta2017detecting} introduced a crowd-sourcing quality assessment approach that is difficult to uncover quality issues automatically. The authors explored most common quality issues in DBpedia datasets, such as incorrect object values, incorrect datatype or language tag and incorrect link. The authors introduced a methodology to adjust crowdsourcing input from two types of audience: \textit{(i)} Linked Data experts through a contest to detect and classify erroneous RDF triples and \textit{(ii)} Crowdsourcing through the Amazon Mechanical Turk. In detail, the authors adapted the Find-Fix-Verify crowdsourcing pattern to exploit the strengths of experts and paid workers. Furthermore, the authors used TripleCheckMate~\cite{kontokostas2013triplecheckmate} a crowdsourcing tool for the evaluation of a large number of individual resources, according to a defined quality problem taxonomy. To understand the quality of data sources, Flemming's~\cite{flemming2010quality} presented an assessment tool that calculates data quality scores based on manual user input for data sources. More specifically, a user needs to answer a series of questions regarding the dataset and assigns weights to the predefined quality metrics. However, it lacks several quality dimensions such as completeness or inconsistency. In~\cite{kontokostas2014test}, Kontokostas \textit{et al.} proposed a methodology for test-driven quality assessment of Linked Data. The authors formalized quality issues and employed SPARQL query templates, which are instantiated into quality test queries. Also, the authors presented RDFUnit\footnote{\url{https://github.com/AKSW/RDFUnit}} a tool centered around schema validation using test-driven quality assessment approach. RDFUnit runs automatically based on a schema and manually generates test cases against an endpoint. RDFUnit has a component that turns RDFS axioms and simple OWL axioms into SPARQL queries that check for data that does not match the axiom.
In contrast, in this study, we aim to learn the constraints (which might not be explicitly stated as RDFS or OWL axioms) as RDF Shapes. Although the overall objectives are similar considering RDF shape induction to this work, for completeness analysis we mainly explore KB evolution analysis. Furthermore, in our approach, we primarily use data profiling information as the input for the process. Results from the consistency analysis can be extended by using RDFUnit for further validation.

\textit{Metadata.} In~\cite{assaf2015roomba}, Assaf \textit{et al.} introduced a framework that handles issues related to incomplete and inconsistent metadata quality. The authors proposed a scalable automatic approach for extracting, validating, correcting and generating descriptive linked dataset profiles. This framework applies several techniques to check the validity of the metadata provided and to generate descriptive and statistical information for a particular dataset or an entire data portal. In particular, the authors extensively used dataset metadata against an aggregated standard set of information. This procedure leads to dependency towards availability of metadata information. Instead, in our approach, we only focus on summary statistics from the collected dataset, and it is independent of external information since the quality profiling can be done only using summary statistics.

\textit{Temporal Analysis.} In \cite{rula2012capturing}, Rula \textit{et al.} started from the premise of dynamicity of Linked Data and focused on the assessment of timeliness in order to reduce errors related to outdated data. A currency metric is introduced to measure timeliness, that is calculated in terms of differences between the observation is done (current time) and the time when the data was modified for the last time. Furthermore, authors also took into account the difference between the time of data observation and the time of data creation. Similarly, in our work, we explore KB dynamicity using data profiling information. Rather than using timeless measures, we investigate the changed behavior present in the dataset using dynamic profiling features introduced by Ellefi \textit{et al.}~\cite{ellefi2017rdf}. 

In~\cite{furber2011swiqa}, Furber and Hepp focused on the assessment of accuracy, which includes both syntactic and semantic accuracy, timeliness, completeness, and uniqueness. One measure of accuracy consists of determining inaccurate values using functional dependence rules, while timeliness is measured with time validity intervals of instances and their expiry dates.  Completeness deals with the assessment of the completeness of schema (representation of ontology elements), completeness of properties (represented by mandatory property and literal value rules), and completeness of population (description of real-world entities). Uniqueness refers to the assessment of redundancy, i.e., of duplicated instances. In this work, we explored the changes present in the KB to identify completeness issues.

Considering the version management and linked data lifecycle, Knuth \textit{et al.}~\cite{knuth2014linked} identified the critical challenges for Linked Data quality. As one of the key factors for Linked Data quality they outlined validation that, in their opinion, has to be an integral part of Linked Data lifecycle. An additional factor for Linked Data quality is version management, which can create problems in provenance and tracking. Finally, as another essential factor they outlined the usage of popular vocabularies or manual creation of new correct vocabularies. Furthermore, Emburi \textit{et al.}~\cite{embury2014feasibility} developed a framework for automatic crawling the Linked Data datasets and improving dataset quality. In their work, the quality is focused on errors in data, and the purpose of the developed framework is to automatically correct errors.

\textit{Statistical analysis.}
Paulheim \textit{et al.}~\cite{paulheim2014improving} presented two approaches SDType and SDValidate for quality assessment. SDType approach help to predict RDF resources type thus completing missing values of rdf:type properties. SDValidate approach detects incorrect links between resources within a dataset. These methods can effectively detect errors on DBpedia; however they require the existence of informative type assertions. Furthermore, more complex errors containing wrong entities with correct types cannot be identified. Taking into account, the probabilistic approach for linked data quality assessment, Li \textit{et al.}~\cite{li2015probabilistic} presented a  probabilistic framework using the relations (equal, greater than, less than) among multiple RDF predicates to detect inconsistencies in numerical and date values based on the statistical distribution of predicates and objects in RDF datasets. However, they mainly focused on identifying errors in the numerical data. In~\cite{ruckhaus2013analyzing}, Ruckhaus \textit{et al.} presented LiQuate, a tool based on probabilistic models to analyze the quality of data and links. The authors used Bayesian Networks and rule-based system for quality assessment. The probabilistic rules are represented by data experts to identify redundant, incomplete and inconsistent links in a set of resources. In our approach, we mainly focus on statistical profiling at the instance level. This reduces the dependency on expert intervention.

In the current state of the art, less focus has been given toward understanding knowledge base resource changes over time to detect anomalies and completeness issues due to the KB evolution. For an evolving KB, we investigated two perspectives: \textit{(i)} Static: data quality analysis with respect to specific tasks without considering dataset dynamics; \textit{(ii)} Dynamic: process of accessing data and temporal analysis, such as timeliness measure.  In Table \ref{tab:qualitySummary}, we summarize the reported linked data quality assessment approaches.

\begin{table*}[!hptb]
	\centering
	\footnotesize
	\tabcolsep=0.11cm
	\caption{Summary of Linked Data Quality Assessment Approaches.} 
	\label{tab:qualitySummary}
	\begin{tabular}{l l p{7cm} l}
		\toprule
		Paper  & Degree of Automation & Goal & Dataset Feature  \\ \midrule

		Bizer \textit{et al.} \cite{bizer2009quality}  & Manual  & WIQA quality assessment framework enables information consumers to apply a wide range of policies to filter information. & Static \\ \addlinespace
		
		Acosta \textit{et al.}~\cite{acosta2017detecting} & Manual & A crowd-sourcing quality assessment approach for quality issues that are difficult to uncover automatically. & Static \\ \addlinespace
		
	    Ruckhaus \textit{et al.}~\cite{ruckhaus2013analyzing}  & Semi-Automatic & LiQuate, a tool based on probabilistic models to analyze the quality of data and links. & Static \\ \addlinespace 
	    
	    Paulheim \textit{et al.}~\cite{paulheim2014improving}   & Semi-Automatic & SDType approach using statistical analysis to predicts classes of RDF resources thus completing missing values of rdf:type properties. & Static \\ \addlinespace 
	    
	    Furber and Hepp~\cite{furber2011swiqa}  & Semi-Automatic  & Focus on the assessment of accuracy, which includes both syntactic and semantic accuracy, timeliness, completeness, and uniqueness. & Dynamics \\ \addlinespace
	    
	    Flemming~\cite{flemming2010quality}  & Semi-Automatic &  Focuses on a number of measures for assessing the quality of Linked Data covering wide-range of different dimensions such as availability, accessibility, scalability, licensing, vocabulary reuse, and multilingualism. & Static \\ \addlinespace 
	 
		Mendes \textit{et al.}~\cite{mendes2012sieve}  & Semi-Automatic &   Sieve framework that uses user configurable quality specification for quality assessment and fusion method. & Dynamic \\ \addlinespace     
		
		Knuth \textit{et al.}~\cite{knuth2014linked}  & Semi-Automatic  &  They outline validation which, in their opinion, has to be an integral part of Linked Data lifecycle. & Static \\ \addlinespace   
		
		Rula \textit{et al.}~\cite{rula2012capturing}  & Automatic  &  Start from the premise of dynamicity of Linked Data and focus on assessment of timeliness in order to reduce errors related to outdated data. & Dynamic \\ \addlinespace 
	    
	    Kontokostas \textit{et al.}~\cite{kontokostas2014test}  & Automatic  &  Propose a methodology for test-driven quality assessment of Linked Data. & Dynamic \\ \addlinespace

		
		Emburi \textit{et al.}~\cite{embury2014feasibility}  & Automatic  &  They developed a framework for automatic crawling the Linked Data datasets and improving dataset quality. &  Dynamic  \\ \addlinespace  
		
		Li \textit{et al.}~\cite{li2015probabilistic}  & Automatic  & They proposed an automatic method to detect error between multi attributes which can not be detected only considering single attribute.  &  Dynamic  \\ \addlinespace   
		
		Assaf \textit{et al.}~\cite{assaf2015roomba}  & Automatic &  They propose a framework that handles issues related to incomplete and inconsistent metadata quality. &  Static \\ \addlinespace 
		
		Debattista \textit{et al.}~\cite{debattista2016luzzu}  & Automatic &  They propose a conceptual methodology for assessing Linked Datasets, proposing Luzzu, a framework for Linked Data Quality Assessment. & Static \\ \addlinespace 
		\bottomrule  \\ 
	\end{tabular}
\end{table*}

\subsection{Knowledge Base Validation}

The problem of knowledge base validation has been explored using \textit{Description Logics} considering both Open World (OW) and Closed World (CW) Assumption. In recent years, various validation languages have been introduced using constraint definitions.

\begin{itemize}

\item The Web Ontology Language (OWL) \cite{mcguinness2004owl} is an expressive ontology language based on Description Logics (DL). The semantics of OWL addresses distributed knowledge representation scenarios where complete knowledge about the domain cannot be assumed. Motik \textit{et al.}~\cite{motik2009bridging} proposed an extension of OWL that attempts to mimic the intuition behind integrity constraints in relational databases. The authors divided axioms into regular and constraints. To address the problem of validation using OWL representation, some approaches use OWL expressions with Closed World Assumption and a weak Unique Name Assumption so that OWL expressions can be used for validation purposes, such as the work presented by Tao \textit{et al.}~\cite{tao2010extending}, and  Stardog ICV\footnote{\url{https://www.stardog.com/docs/}}. 

\item The Shape Expressions (ShEx)~\cite{shex} language describes RDF nodes and graph structures. A node constraint describes an RDF node (IRI, blank node or literal) and a shape describes the triples involving nodes in an RDF graph. These descriptions identify predicates and their associated cardinalities and datatypes.

\item The W3C Shapes Constraint Language (SHACL)~\cite{shacl} is used for validating RDF graphs against a set of conditions. These conditions are provided as shapes and other constructs expressed in the form of an RDF graph. In particular, it helps to identify constraints using SPARQL. Also, it provides a high level vocabulary to identify predicates and their associated cardinalities, and datatypes.
SHACL is divided into two parts: \textit{(i)} SHACL Core, describes a core RDF vocabulary to define common shapes and constraints; and \textit{(ii)} extension mechanism named SHACL-SPARQL. In this work, we explore the SHACL Core for consistency evaluation. We look at SHACL Shape for a specific class to identify constraints components. In SHACL, a \textit{Shape} is defined as the collection of targets and constraints components. Targets specify which nodes in the data graph must conform to a shape and constraint components determine how to validate a node. \textit{Shapes graph} represent an RDF graph that contains shapes. Conversely, \textit{Data graph} represents an RDF graph that contains data to be validated. Furthermore, SHACL defines two types of Shapes: \textit{(i)} \textit{Node shapes} presents the constraints information about a given focus node; and \textit{(ii) Property shapes} present constraints about a property and values of a path for a node.

\item SPARQL Inference Notation (SPIN)\footnote{\url{http://spinrdf.org}} constraints associate RDF types or nodes with validation rules. In particular, it allows users to use SPARQL to specify rules and logical constraints. 

\end{itemize}

These shape expression languages, namely, ShEx, SHACL, and SPIN, aim to validate RDF data and to communicate data semantics among users. They cover constraints such as keys and cardinality; however, their expressivity is limited and require user interventions in every step. 
Furthermore, various research endeavors explored the RDF validation based on the Closed World Assumption (CWA). For example, Patel-Schneider~\cite{Patel-Schneider:2015:UDL:2887007.2887042} explored Description Logics as a mean to provide the necessary framework for checking constraints and providing facilities to analyze CWAs. The authors utilized inference as a mean for constraint checking, which is the core service provided by Description Logics. 
Our final goal is different from these research approaches. In particular, we study how data profiling can be applied to constraints based feature extraction in a predictive setting.  
For example, cardinality estimation has been studied in many different domains including relational data. In addition, integrity constraints for validation tasks has many other applications, such as network monitoring for detecting DDoS attacks or worm propagation, link based spam detection, and relation join query optimization. The existing cardinality estimation algorithms such as Hit Counting~\cite{flajolet1985probabilistic}, Adaptive Sampling~\cite{flajolet1990adaptive}, Probabilistic Counting~\cite{whang1990linear} and HYPERLOGLOG~\cite{heule2013hyperloglog} aim to estimate the number of distinct elements in very large datasets with duplicate elements. For cardinality estimation in RDF data, Neuman and Moerkotte~\cite{neumann2011characteristic} have proposed ``characteristic sets'' for performing cardinality estimations using SPARQL queries with multiple joins. Overall, these works differ from the work presented in this paper on two axes. First, they are focused on determining the cardinalities of each value rather than the cardinality of the entity-value relation. Second, they are focused on query optimization rather than integrity constraint validation. 
We consider the profiling of instances as a mean to estimate constraint values which can help to understand consistency issues. 



\section{Completeness and Consistency Analysis}\label{sec:metrics}

In this section, we investigate the concept of KB evolution analysis to derive completeness measurement functions. 
For consistency analysis, we explore integrity constraints using shape induction for feature extraction.

\subsection{Evolution Analysis and Dynamic Features}

Large KBs are often maintained by communities that act as curators to ensure their quality~\cite{dbpediaChanges2016}. The benefit of KB evolution analysis is two-fold~\cite{meimaris2015framework}: (1) quality control and maintenance; and (2) data exploitation. Considering quality control and maintenance, KB evolution can help to identify common issues such as broken links or URI changes that create inconsistencies in the dataset. On the contrary, data exploitation can provide valuable insights regarding dynamics of the data, domains, and the communities that explore operational aspects of evolution analysis~\cite{meimaris2015framework}. KBs naturally evolve due to several causes: \textit{(i)} resource representations and links that are created, updated, and removed; \textit{(ii)} the entire graph can change or disappear. 
The kind of evolution that a KB is subjected to depends on several factors, such as:
\begin{itemize}
\item \textit{Frequency of update:} KBs can be updated almost continuously (e.g. daily or weekly) or at long intervals (e.g. yearly);
\item \textit{Domain area:} depending on the specific domain, updates can be minor or substantial. For instance, social data is likely to experience wide fluctuations than encyclopedic data, which is likely to undergo smaller knowledge increments;
\item \textit{Data acquisition:} the process used to acquire the data to be stored in a KB and the characteristics of the sources may influence the evolution. For example, updates on individual resources cause minor changes when compared to a complete reorganization of a data source infrastructure such as a change of the domain name;
\item \textit{Link between data sources:} when multiple sources are used for building a KB, the alignment and compatibility of such sources affect the overall KB evolution. The differences of KBs have been proved to play a crucial role in various curation tasks such as the synchronization of autonomously developed KB versions, or the visualization of the evolution history of a KB~\cite{papavasileiou2013high} for more user-friendly change management.
\end{itemize}

Ellefi~\textit{et al.}~\cite{ellefi2017rdf} presented a set of dynamic features for data profiling. In this work, we explore these dynamic features for measuring the completeness quality characteristics. Based on Ellefi~\textit{et al.}~\cite{ellefi2017rdf}, we explored the following dynamic features: 

\begin{itemize}

\item \textit{Lifespan}: knowledge bases contain information about different real-world objects or concepts commonly referred as entities. Lifespan represents the period when a certain entity is available and it measures the change patterns of a knowledge base. Change patterns help to understand the existence and the categories of updates or change behavior.

\item \textit{Degree of change:} it helps to understand to what extent the performed update impacts the overall state of the knowledge base. Furthermore, the degree of changes helps to understand what are the causes for change triggers as well as the propagation effects.

\item \textit{Update history}: it contains basic measurement elements regarding the knowledge base update behavior such as frequency of change. The frequency of change measures the update frequency of KB resources. For example, the instance count of an entity type for various versions. 

\end{itemize}

\subsection{Completeness Analysis based on Dynamic Features}\label{sec:completeness}

The ISO/IEC 25012 standard~\cite{SQUARE} refers to completeness as the degree to which subject data associated with an entity has values for all expected attributes and related entity instances in a specific context of use.
In this paper, for completeness characteristics, 
we look into the dynamic features using periodic data profiling in order to identify quality issues. 
Taking into account linked data dynamics,\footnote{\url{https://www.w3.org/wiki/DatasetDynamics}} the update behaviour of classes and properties can be stable/growth or unstable~\cite{meimaris2015framework}. 
Table~\ref{tab:propCompleteness} illustrates two common types of change behaviour using property frequency as measurement element.

\begin{table}[!hptb]
\centering
\small
\caption{Categories of change behaviour.}
\label{tab:propCompleteness}
\begin{tabular}{lp{5cm}}
\toprule
\textbf{Type} & \textbf{Description} \\ \midrule
Stable/Growth = 1  & If the property frequency at release $N$ equal or greater than $N-1$ \\ \addlinespace
Unstable = 0  & If the property frequency at release $N$ less than $N-1$\\ \bottomrule
\end{tabular}
\end{table}

\subsubsection{Measurement Elements}

Statistical operations using data profiling provides descriptive information about data types and patterns in the dataset. For example, property distributions, number of entities, and number of predicates. For computing the change detection, we used basic statistical operations. We thereby use the following key statistics: 
\textit{(i)} number of distinct predicates;
\textit{(ii)} number of distinct subjects; 
\textit{(iii)} number of distinct entities per class; 
\textit{(iv)} frequency of predicates per entity. 

In particular, we aim to detect variations of two basic statistical measures that can be evaluated with the most simple and computationally inexpensive operation, i.e., counting. 

The computation is performed on the basis of the classes in a KB release of $V$, i.e. given a class $C$ we consider all entities $E$ of the type $C$ as:


$$\text{count}( C ) = | \{ s : \exists \langle s, \text{typeof}, C \rangle \in V \} |$$

The $\text{count}(C)$ measurement can be performed with a SPARQL query such as:

\begin{lstlisting}[language=SQL,basicstyle=\footnotesize\ttfamily]
SELECT COUNT(DISTINCT ?s) AS ?COUNT
WHERE {  ?s a <C> . }
\end{lstlisting}

The second measure element focuses on the frequency of the properties, within a class $C$. We define the frequency of a property (in the scope of class $C$) as:

$$\text{freq}(p, C) = | \{ \langle s, p, o \rangle : \exists \langle s, p, o \rangle  \wedge \langle s, \text{typeof}, C \rangle \in V \} | $$


The $\text{freq}(p,C)$ measurement can be performed with a SPARQL query having the following structure:

\begin{lstlisting}[language=SQL,basicstyle=\footnotesize\ttfamily]
SELECT COUNT(*) AS ?FREQ
WHERE {  
    ?s <p> ?o.
    ?s a <C>.
}
\end{lstlisting}

There is an additional basic measure element that can be used to build derived measures: the number of properties present for the entity type $C$ in the release $i$ of the KB. Therefore, distinct property count of entity type $C$ as:

$$\textit{NP}(C) = | \{ p : \exists \langle s, p, o \rangle  \wedge \langle s, \text{typeof}, C \rangle \in V \} | $$

The $\textit{NP}(C)$ measure can be collected with a SPARQL query having the following structure:

\begin{lstlisting}[language=SQL,basicstyle=\footnotesize\ttfamily]
SELECT COUNT(DISTINCT ?p) AS ?NP
WHERE {  
    ?s ?p ?o.
    ?s a <C>.
}
\end{lstlisting}

The essence of the proposed approach is the comparison of the measure across distinct releases of a KB. In the remainder, we will use a subscript to indicate the release that the measure refers to. The releases are numbered progressively as integers starting from one and, by convention, the most recent release is $n$.
So that, for instance, $count_{n-1}( \textit{foaf:Person} )$ represents the count of resources typed with \emph{foaf:Person} in the last but one release of the knowledge base under consideration.
More specifically we used the property frequency for a specific class for completeness metrics. 

\subsubsection{Measurement Functions} \label{sec:completenessMeasure}

On the basis of the dynamic features~\cite{ellefi2017rdf}, a further conjecture drive that the growth of knowledge in a mature KB ought to be stable. Furthermore, we argue that completeness issues can be identified through monitoring lifespan of an RDF KBs.
A simple interpretation of the stability of a KB is monitoring the dynamics of knowledge base changes~\cite{rifat2018}. This measure could be useful to understand high-level changes by analyzing KB growth patterns.

We can monitor growth level of KB resources (instances) by measuring changes presented in different releases. In particular, knowledge base growth can be measured by detecting the changes over KB releases utilizing trend analysis such as the use of simple linear regression. Based on the comparison between observed and predicted values, we can detect the trend in the KB resources, thus detecting anomalies over KB releases if the resources have a downward trend over the releases.

We derive KB lifespan analysis regarding change patterns over time. To measure the KB growth, we applied linear regression analysis of entity counts over KB releases. 
In the regression analysis, we checked the latest release to measure the normalized distance between an actual and a predicted value. 
In particular, in the linear regression we used entity count ($y_{i}$) as dependent variable and time period ($t_{i}$) as independent variable. 
Here, $n=$ \textit{total number of KB releases} and $i=1...n$ present as the time period.

We start with a linear regression fitting the count measure of the class (C):
$$y(C)=a \cdot t+b$$
The residual can be defined as:
$$residual_i(C) = a \cdot t_i + b - count_i(C)$$

We define the normalized distance as:
\[ND(C)=\frac{residual_n(C)}{mean(|residual_{i}(C)|)}\]

Based on the normalized distance, we can classify the growth of a class C as:

\[
Growth (C) =\left\{
\begin{array}{ll}
1 & if (ND(C)) \geq 1  \\

0 & if (ND(C)) < 1
\end{array}
\right.
\]

The value is 1 if the normalized distance between actual value is higher than the predicted value of type $C$, otherwise it is 0. In particular, if the KB growth prediction has the value of 1 then the KB may have an unexpected growth with unwanted entities otherwise the KB remains stable.

To further validate our assumptions, we explore the completeness of properties by monitoring the variations due to KB updates. More specifically, by property completeness analysis we focus on the removal of information as an adverse effect of the KB evolution. We can use the frequency of predicates of an entity type as the essential measurement element. Furthermore, by comparing property frequency between two KB releases, we can detect completeness issues.
Considering the changed behavior presented in Table~\ref{tab:propCompleteness}, the value of 0 means that a property presents in the last release might have completeness issues. 

The basic measure we use is the frequency of predicates, in particular, since the variation in the number of subjects can affect the frequency, we introduce a normalized frequency as:

$$
\text{NF}_i(p,C) = \frac{\text{freq}_i(p,C)}{\text{count}_i(C)}
$$

On the basis of this derived measure we can thus define completeness of a property $p$ in the scope of a class $C$ as:

$$
\textit{\small Completeness}_i(p,C)=
\begin{cases}
1 ,& \text{NF}_i(p,C)\geq\text{NF}_{i-1}(p,C)\\
0 ,& \text{NF}_i(p,C)<\text{NF}_{i-1}(p,C)     
\end{cases}
$$

where $\textit{NF}_i(C)$ is the number of properties present for class $C$ in the release $i$ of the knowledge base.

At the class level the completeness is the proportion of complete predicates and can be computed as:
$$
\textit{Completeness}_i(C) = \frac{\sum\limits_{k=1}^{\text{NP}_i(C)} \textit{Completeness}_i(p_k,C)}{\text{NP}_i(C)}
$$

\subsection{Consistency Analysis based on Integrity Constraints}\label{sec:integrityConstraints}

Consistency checks whether inconsistent facts are included in the KB~\cite{zaveri2016quality}. 
For accessing consistency, we can use an inference engine or a reasoner, which supports the expressivity of the underlying knowledge representation formalism. In the context of KB validation, languages, such as W3C Shapes Constraint Language (SHACL) and Shape Expressions Language (ShEx), allow integrity constraints to be defined for validation tasks.
In this work, we explore the integrity constraints definitions present in SHACL core for consistency evaluation.  We generate shapes at the class-level using data profiling information. We consider three constraints for consistency checks for evolving KBs: cardinality, range, and string constraint. We consider these three constraints based on the following conditions: \textit{(i)} to evaluate properties with correct specifications, we explore cardinality constraints to identify the correct mapping of properties for a specific class, and \textit{(ii)} to evaluate contradictions within the data, we explore the range constraint values.

Taking into account profiling based shape induction tasks, we compute the RDF term at instance-level using the data instances only. We thereby use the following key statistics: 
\textit{(i)} percentage (\%) of IRIs, blank nodes, and literals;
\textit{(ii)} no. of triples with IRI and its frequency, length, namespace, patterns;
\textit{(iii)} no. of triples with Literals (String/Numeric/Dates) and its frequency, language, length, patterns, min, max, mean, std, variance.

The motivation for using these key statistics is that these statistics could provide some insights related to different possible distributions to identify feature vectors. The percentage (\%) of IRIs, blank nodes, and literals are used to extract Range constraints. Also, no. of triples with IRI and its frequency, length, namespace, patterns is used for Range contraints feature extraction. For cardinality and string constraints feature extraction, we considered the triples with literals (String/Numeric/Date) and its frequency, language, length, patterns, min, max, mean, std, variance. For example, based on the raw cardinality value distributions, we can compute the distinct cardinality values. Overall, we derive 11 statistical measures including min-max cardinalities, mean, mode, standard deviation, variance, quadratic mean, skewness, percentiles, and kurtosis~\cite{freedman2009statistical}. Our intuition is that these values are descriptive to classify the constraints category. Nevertheless, the data can be noisy, and either min or/and max could be outliers. To address this, we add statistical features that give more insights about the distribution of the cardinalities such as mean, mode, kurtosis, standard deviation, skewness, variance and four percentiles.

In the remainder of this section, we describe cardinality, range, and string constraints for feature extraction process. We describe each constraint with examples based on the English DBpedia 201604 release.

\begin{description}

\item[Cardinality constraints.] We observe a trend in vocabularies where the cardinality constraints are explicitly expressed~\cite{nandanardf}. When, we analyzed the 551 vocabularies in the \textit{Linked Open Vocabularies (LOV)} catalogue for the values of owl:minCardinality, 96.91\% (848 out of 875) of owl:maxCardinality constraints have value 1 and 93.76\% (631 out of 673) of the owl:minCardinality values are either 0 or 1~\cite{nandanardf}. 
Thus, in this work, we explore which cardinality category each property has with respect to a given class. By doing so, we present cardinality value estimation as a classification problem.
Table~\ref{tab:cardinalityPatterns} shows the common cardinality patterns.

For the classification task, we use the five main types of cardinality classes: MIN0, MIN1, MIN1+, MAX1, and MAX1+. Out of these, MIN0 and MAX1+ do not put any constraints on the data, such that, any data will be valid for those cardinality types. Thus, if we detect those types, we do not generate constraints. For other types, corresponding SHACL property constraints are generated as illustrated in Listing~\ref{lst:cardcons}.

\begin{lstlisting}[caption={Cardinality constraints. },label={lst:cardcons}]
@prefix dbo: <http://dbpedia.org/ontology/> .
@prefix sh: <http://www.w3.org/ns/shacl#> .

ex:DBpediaPerson a sh:NodeShape;
 sh:targetClass dbo:Person;   
# for MIN1 and MIN1+ 
 sh:property [sh:path foaf:name; 
  sh:minCount 1 ];
  
# for MAX1  
 sh:property [ sh:path dbo:birthDate;
  sh:maxCount 1] .
 
# for MAX1+   
 sh:property [ sh:path dbo:union;
  sh:maxCount 1] .
  
\end{lstlisting}

\begin{table}[!hptb]
\centering
\small
\caption{Minimum and maximum cardinality levels.}
\label{tab:cardinalityPatterns}
\begin{tabular}{ll}
\toprule
\textbf{Key} & \textbf{Description} \\ 
\midrule
MIN0  & Minimum Cardinality = 0            \\ \addlinespace
MIN1  & Minimum Cardinality = 1            \\ \addlinespace
MIN1+ & Minimum Cardinality \textgreater 1 \\ \addlinespace
MAX1  & Maximum Cardinality = 1            \\ \addlinespace
MAX1+ & Maximum Cardinality \textgreater 1 \\ \addlinespace
\bottomrule
\end{tabular}
\end{table}
    
We generate cardinality information for each property associated with the instances of a given class. 
The work presented by Neumann and Moerkotte~\cite{neumann2011characteristic} helps to identify raw cardinality values using SPARQL queries. They proposed a highly accurate cardinality estimation method for RDF data using star joins SPARQL queries.
Similarly, we also explore the process of cardinality values estimation using the results from SPARQL queries. Thus, our cardinality constraints generation process is based on the study presented by Neumann and Moerkotte~\cite{neumann2011characteristic}. 
We collected distinct cardinality values by using star join SPARQL query. An example of join SPARQL queries for raw cardinality values estimation is presented in Listing~\ref{lst:card}.
    
\begin{lstlisting}[language=SQL,caption={SPARQL query for the cardinality value estimation.},label={lst:card},basicstyle=\footnotesize\ttfamily]
    SELECT ?card (COUNT (?s) as ?count )
    WHERE {
      SELECT ?s (COUNT (?o) as ?card) 
      WHERE {
         ?s a ?class ;
         ?p ?o
       } GROUP BY ?s
    } GROUP BY ?card ORDER BY DESC(?count)
    
\end{lstlisting}

\item[Range constraints.] We use the subset of the target Node already identified in SHACL, i.e., \textit{IRI}, \textit{Literal}, \textit{BlankNode}, and \textit{BlankNodeOrIRI}. Table \ref{tab:ObjectType} illustrates the target Node objects type in SHACL. Each value of target Node in shape is either an IRI or a literal. For range constraints, our goals are twofold. First, we want to generate an object as target node constraint for each property associated with a given class. Once the target node type is determined, then more specific range constraints have to be decided. If the node type is \textit{Literal}, the corresponding datatype has to be determined. If the node type is either \textit{IRI}, \textit{BlankNode}, or \textit{BlankNodeOrIRI} the class type of the objects has to be determined. 

\begin{table}[!hptb]
\centering
\small
\caption{Objects Type.}
\label{tab:ObjectType}
\begin{tabular}{cccc}
\toprule
\textbf{IRI} & \textbf{BlankNode} & \textbf{Literal} & \textbf{Type} \\ 
\midrule
X & X & X & Any \\ \addlinespace
X & X &  & BlankNodeOrIRI \\ \addlinespace
X &  &  & \textbf{IRI} \\ \addlinespace
 & X &  & BlankNode \\ \addlinespace
 &  & X & \textbf{Literal} \\ \addlinespace
X &  & X & IRIOrLiteral \\ \addlinespace
 & X & X & BlankNodeOrLiteral \\ \addlinespace

\bottomrule
\end{tabular}
\end{table}

We classify each property associated with instances of a given class to one of the aforementioned node types. The second task of assigning the corresponding datatype or class as the range of each property is done based on heuristics of datatype or class type distributions among the set of objects associated with the property. 
For example, \textit{dbo:Web} has distribution of 73.13\% for IRI node type and 26.89\% for LIT node type for \textit{dbo:SoprtsTeam} entity type. In this account, IRI has larger distribution then LIT node type. Based on our heuristics, we considered \textit{dbo:Web} node type as IRI. An example of \textit{dbo:Person}-\textit{dbp:birthPlace} objects nodeKind constraints Shape is illustrated in the Listing \ref{lst:nodetype}.

\begin{lstlisting}[caption={Node type constraints. },label={lst:nodetype}]
@prefix dbo: <http://dbpedia.org/ontology/> .
@prefix dbp: <http://dbpedia.org/property/> .
@prefix sh: <http://www.w3.org/ns/shacl#> .

ex:DBpediaPerson a sh:NodeShape;
 sh:targetClass dbo:Person;   
# node type IRI
 sh:property [sh:path dbp:birthPlace; 
  sh:nodeKind sh:IRI;
  sh:or ( [sh:class schema:Place] 
    [ sh:class dbo:Place ] )
  ];
  
# node type literal
 sh:property [ sh:path dbp:deathDate;
  sh:nodeKind sh:Literal;
  sh:datatype xsd:date ] .
\end{lstlisting}

\item[String based Constraints.] For string based constraints generation the primary focus is to understand minimum length (minLength) and maximum length (maxLength) of a property. In this context max and min length subjected to rdf:type and node with literal values. In general, if the value of minLength is 0, then there is no restriction on the string length, but the constraint is still violated if the value node is a blank node.
On the other hand, the value of maxLength without restriction could be any string length based on the rdf:type. 
We considered the distribution of string lengths to identify minLength and maxLength of literal values of a property. More specifically, we explored all the properties present in a class, and interquartile range of string literals lengths distribution for constraints generation. We evaluate the minLength using 1st quartile(Q1) and maxLength using the 3rd quartile (Q3). Table \ref{tab:stringLengthPatterns} illustrates the string length conditions for minLength and maxLength.
In particular, we mainly focus on identifying a relative range for the maximum and minimum length. An example of string length based SHACL Shape for \textit{dbo:title} property is presented in Listing~\ref{lst:string}.

\begin{table}[!hptb]
\centering
\small
\caption{Minimum and maximum String length levels.}
\label{tab:stringLengthPatterns}
\begin{tabular}{ll}
\toprule
\textbf{Key} & \textbf{Description} \\ 
\midrule
minLength0  & Minimum Length \textless  Q1            \\ \addlinespace
minLength1  & Minimum Length $\geq$ Q1            \\ \addlinespace
maxLength0 & Maximum Length \textless Q3            \\ \addlinespace
maxLength1 & Maximum Length $\geq$  Q3 \\ \addlinespace
\bottomrule
\end{tabular}
\end{table}

\begin{lstlisting}[caption={String constraints. },label={lst:string}]
@prefix dbo: <http://dbpedia.org/ontology/>.
@prefix sh: <http://www.w3.org/ns/shacl#>.

ex:DBpediaPerson a sh:NodeShape;
 sh:targetClass dbo:Person;   
# minLength 
 sh:property [sh:path foaf:name; 
  sh:minLength 1; 
  sh:maxLength 8];
  
# for MAX1  
 sh:property [ sh:path dbo:birthDate;
  sh:minLength 1;
  sh:maxLength 8] .
\end{lstlisting}

\end{description}

\section{Approach}
\label{sec:approach}

%

\begin{figure*}[!ht]
    \centering
    \includegraphics[width=.8\linewidth]{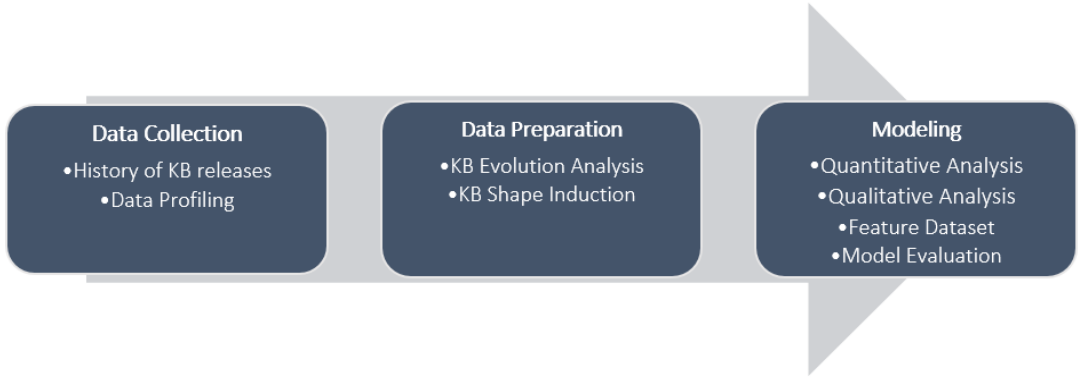}
    \caption{Process flow of the proposed completeness and consistency analyses.}
    \label{fig:overall}
\end{figure*}

In order to formulate an answer to the research questions, an approach is designed to identify KB with completeness and consistency issues.
Based on the data profiling information, a set of features is introduced taking into account completeness and consistency analysis. These features are applied in the learning models to create RDF shapes. Figure \ref{fig:overall} illustrates the process flow of the proposed approach that is divided in three main stages:

\textit{(i)} \textit{Data Collection:} A data curator needs to select an entity type to initiate the completeness analysis procedure.
Then, the process checks the chosen entity types present in all KB releases to verify schema consistency and computes the summary statistics.

\textit{(ii)} \textit{Data Preparation:} The features are generated using the results from data profiling.

\textit{(iii) Modeling:} The validation of the hypothesis is performed using quantitative and qualitative analysis. Also, the performance of the constraints classifiers is assessed using learning models. 

\begin{figure*}[!ht]
	\centering
	\includegraphics[width=\linewidth]{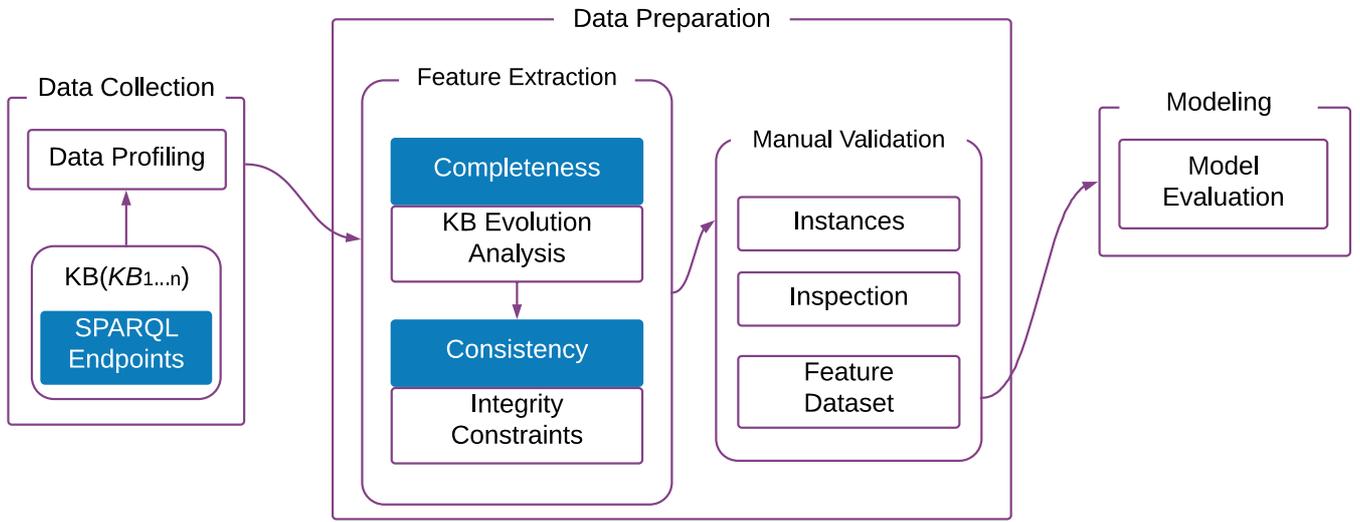}
	\caption{Workflow of the completeness and consistency analyses.}
	\label{fig:workflow}
\end{figure*}

Figure~\ref{fig:workflow} illustrates the completeness and consistency analysis workflow that is outlined in Figure \ref{fig:overall}. The stages are explained in details below.

\subsection{Data Collection}

In this approach, the history of KB releases and summary statistics are applied as inputs to the completeness and consistency analysis.
The acquisition of KB releases is performed by querying multiple SPARQL endpoints (assuming each release of the KB is accessible through a different endpoint) or by loading data dumps. For each KB releases, summary statistics are generated using data profiling. 
The Data Collection component is built on top of Loupe~\cite{loupe}, an online system that inspects and extracts automatically statistics about the entities, vocabularies (classes, and properties), and frequent triple patterns of a KB. 

In this component, preprocessing operations is perfomed over the collected dataset based on schema consistency checks. 
It is essential to perform the schema consistency checks due to high-level changes are more schema-specific and dependent on the semantics of data. Hence, this component does the following tasks: \textit{(i)} selection of only those entity types that are present in all KB releases, and \textit{(ii)} for each entity type, selection of only those properties present in that class. For example, in the implementation, the properties are filtered for an entity type in case the instance count is 0 for all the KB releases.  

\subsection{Data Preparation}

The goal of this stage is to extract the completeness and consistency features for quantitative and qualitative analysis. These automatically generated features are further validated using manual validation (which is a human-driven task). The data preparation process is divided into two stages: \textit{(i)} feature extraction; and \textit{(ii)} manual validation. 
The feature extraction component is based on evolution based completeness analysis, and integrity constraints based consistency analysis.
The entity types with completeness issues are considered as input to the constraints based feature extraction process. Then, this feature dataset is used for integrity constraints based evaluation tasks. 
Furthermore, qualitative analysis is performed using manual validation to evaluate the precision of completeness measure. Also, the features are manually evaluated to create a partial gold standard. The components are explained in detail below.

\subsubsection{Feature Extraction}\label{FeatureEngineering}

A feature extraction task is performed to instruct the learning models. It is composed of two stages: \textit{(i)} selecting an entity type using the completeness measure results, and \textit{(ii)} constraints based shape induction to compute the features. The features are four in total and they are grouped into two categories as shown in Table~\ref{tab:feature}. 

\begin{table}[!hptb]
 \centering
  \small
   \caption{Features based on quality issues.}
    \label{tab:feature}
   \begin{tabular} {lll}
   \toprule
  \textbf{ Quality Issues} & \textbf{Feature} & \textbf{Classifier Values}\\ 
  \midrule
    Completeness &   Property & (0,1) \\\addlinespace
    Consistency     & Minmum Cardinality  & (MIN0,MIN1+) \\\addlinespace
        & Maximum Cardinality  & (MAX1, MAX1+) \\\addlinespace
                     & Range & (IRI,LIT) \\  \addlinespace
    \bottomrule
    \end{tabular}
    \end{table}

\textbf{\textit{i)} \textit{Completeness features:}} These features are extracted using evolution based completeness analysis~\ref{sec:completeness}). In particular, using the selected class and properties from schema consistency check, completeness is measured by comparing the changes present in the current release with respect to the previous release. 
The result of the completeness features is indicated by a Boolean value $0$ or $1$: $1$ indicates a normal growth, while $0$ indicates an unstable behaviour (Table~\ref{tab:propCompleteness}).


\textbf{\textit{ii)} \textit{Consistency features:}} These features are based on the results from integrity constraint checks that are derived from the SHACL representation (Section~\ref{sec:integrityConstraints}). 
In particular, the experimental analysis is based on cardinality, and range constraints. Moreover, the cardinality constraints is divided into minimum and maximum cardinality constraints. In particular, this phase evaluate the constraints based feature dataset using three constraints: \textit{i)} Properties with minimum cardinality values of MIN0 or MIN1+; \textit{ii)} Properties with maximum cardinality values of MAX1 or MAX1+; \textit{iii)} Properties with range values of LIT or IRI. Finally, each of this feature is used as inputs and applied in supervised learning models to evaluate the constraints based classifier performance.

\subsubsection{Manual validation and gold standard creation}\label{sec:ManualValidation}

The main goal of this step is to extract, inspect, and perform manual validation to identify the causes of quality issues as well as create gold standard. Manual validation tasks are based on the following three steps:

\textit{i)} \textit{Instances}: For manual validation, a portion of the properties with quality issues is selected using the completeness analysis. The selection is performed in a random fashion to preserve the representativeness of the experimental data. 
The proposed completeness and consistency analysis is based on the results of statistical data profiling to identify any missing entities. 
Based on the quantitative analysis results, in this step, all entities are extracted from the two releases of a given KB and set disjoint operation is performed to identify missing entities.  

\textit{ii)} \textit{Inspections}: Using the dataset from instance extraction phase, an inspection of each entity is performed for manual validation and report. Various KBs adopt automatic approaches to gather data from the structured or unstructured data sources. For example, the DBpedia KB uses an automatic extraction process based on the mapping with Wikipedia pages. For the manual validation, a source inspection is performed using the missing instances to identify the causes of quality issues. In particular, a manual evaluation checks if the information is present in the data sources but missing in the KB.

\textit{iii)} \textit{Report}: the validation result of an entity is reported as true positive (the subject presents an issue, and an actual problem was detected) or false positive (the item presents a possible issue, but none actual problem is found).

In this approach, a partial gold standard strategy (Sec.~\ref{sec:goldStandard}) is adopted based on the assumption that a new (small) training set is needed when dealing with a new knowledge base. The manual validation phase is then in charge of inspecting and performing a manual annotation of the detected integrity constraints. In detail:

\textit{(i) Feature extraction}: At first, the entities and properties are selected from the completeness analysis results for constraints based feature extraction. Then, it selects the properties annotated with integrity constraints for further inspection. 

\textit{(ii) Inspection}: the validation result of an instance is reported as \textit{Correct} (the properties are annotated with correct integrity constraint) or \textit{Incorrect} (the item presents a wrong integrity constraint). 

\textit{(ii) Feature dataset}: the outcome of the manual validation tasks is a subset of the feature dataset according to each integrity constraints. This dataset is considered as the training set for the modeling phase.

\subsection{Modeling}

In this phase, five learning models are applied to evaluate the performance of the cardinality and range constraints classifier by computing precision, recall, and F-measure (Sec.~\ref{sec:ML}).
The modeling task is run with a \textit{10-fold cross validation} setup in standard settings. The performance is measured using five classical learning models (Section~\ref{sec:ML}). These models are selected to evaluate classifiers performance considering the diversity in machine learning algorithms and to identify the best performing model. Based on the empirical analysis the best performing model is applied for the prediction tasks.

\section{Experiments and Evaluations}\label{sec:experimentalAnalysis}

This section describes the experiments performed on two KBs, namely DBpedia (both English and Spanish versions) and 3cixty Nice. We first present the experimental setting of the implementation, and then, we report the results of both \textit{(i)} completeness analysis based on dynamic features and \textit{(ii)} consistency analysis using integrity constraints.

\subsection{Experimental Settings}

We selected 3cixty Nice KB and DBpedia KB according to: \textit{(i)} popularity and representativeness in their domain: DBpedia for the encyclopedic domain, 3cixty Nice for the tourist and cultural domain; \textit{(ii)} heterogeneity in terms of content being hosted such as periodic extraction of various event information collected in 3cixty Nice KB, \textit{(iii)} diversity in the update strategy: incremental and usually as batch for DBpedia, continuous update for 3cixty. In detail: 

\begin{itemize}
	\item \textit{3cixty Nice} is a knowledge base describing cultural and tourist information concerning the cities of Nice. This knowledge base was initially developed within the 3cixty project,\footnote{\url{https://www.3cixty.com}} which aimed to develop a semantic web platform to build real-world and comprehensive knowledge bases in the domain of culture and tourism for cities. The KB contains descriptions of events and activities, places and sights, transportation facilities as well as social activities, collected from local and global data providers, and social media platforms.
	
	\item \textit{DBpedia}\footnote{\url{http://wiki.dbpedia.org}} is among the most popular knowledge bases in the LOD cloud. This knowledge base is the output of the DBpedia project that was initiated by researchers from the Free University of Berlin and the University of Leipzig, in collaboration with OpenLink Software. DBpedia is roughly updated every year since the first public release in 2007. DBpedia is created from automatically extracted information contained in Wikipedia,\footnote{\url{https://www.wikipedia.org}} such as infobox tables categorization information, geo-coordinates, and external links. 	
\end{itemize}

Following we present a detailed summary of the extracted datasets for each KB.

\textit{3cixty Nice.} In the data collection module, we used the private SPARQL endpoint for the 3cixty Nice KB. We collected two datasets: \textit{(i)} 8 snapshots based on each 3cixty Nice release, and \textit{(ii)} daily snapshots over the period of 2 months.  
The 3cixty Nice KB schema~\cite{3cixty} remained unchanged for all eight releases collected from 2016-06-15 to 2016-09-09. 
We collected those instances having the \textit{rdf:type} of \textit{lode:Event} and \textit{dul:Place}. The distinct entity count for \textit{lode:Event} and \textit{dul:Place} is presented in Table~\ref{tab:EntityCount}. Overall, we collected a total of $149$ distinct properties for the \textit{lode:Event} typed entities and $192$ distinct properties for the \textit{dul:Place} typed entities across eight different releases. 
Furthermore, to monitor the completeness issues for continuous updates, we collected 50 snapshots of \textit{lode:Event} entity type from 2017-07-27 to 2017-09-16. The daily snapshots values are collected without considering distinct count to investigate the changes present in the data extraction pipeline. Table~\ref{tab:lodeEvent} reports the entity count of \textit{lode:Event} type using periodic snapshots generation. 

\begin{table}[!ht]
    \centering
    \small
\caption{Distinct entity count of \textit{lode:Event} and \textit{dul:Place} types.}
\label{tab:EntityCount}
\begin{tabular}{ccc}
\toprule
\textbf{Release} & \textbf{lode:Event} & \textbf{dul:Places} \\ \midrule
	2016-03-11 & 605 & 20,692 \\ \addlinespace
	2016-03-22 & 605 & 20,692\\ \addlinespace
2016-04-09 & 1,301  & 27,858  \\ \addlinespace
2016-05-03  & 1,301  & 26,066 \\ \addlinespace
	2016-05-13  & 1,409  & 26,827  \\ \addlinespace
2016-05-27  & 1,883  & 25,828 \\ \addlinespace
2016-06-15  & 2,182  & 41,018  \\ \addlinespace
2016-09-09 & 689  & 44,968 \\ 
\bottomrule
\end{tabular}
\end{table}

\begin{table}[!ht]
    \centering
    \small
\caption{Periodic snapshots of \textit{lode:Event} class.}
\label{tab:lodeEvent}
\begin{tabular}{cc}
\toprule
\textbf{Release} & \textbf{Entity Count} \\ 
\midrule
2017-07-27 & 114,054 \\ \addlinespace
2017-07-28 & 114,542 \\ \addlinespace
2017-07-29 & 114,544 \\ \addlinespace
2017-07-30 & 114,544 \\ 
\midrule
\multicolumn{2}{c}{other rows are omitted for brevity} \\ 
\midrule
2017-09-14 & 188,967 \\ \addlinespace
2017-09-15 & 192,116 \\ \addlinespace
2017-09-16 & 154,745 \\ \addlinespace
\bottomrule
\end{tabular}
\end{table}

\textit{DBpedia.} We collected a total of $11$ DBpedia releases from which we extracted $4477$ unique properties. For this analysis we considered the following ten classes: \emph{dbo:Animal}, \emph{dbo:Artist}, \emph{dbo:Athlete}, \emph{dbo:Film},  \emph{dbo:MusicalWork}, \emph{dbo:Organisation}, \emph{dbo:Place}, \emph{dbo:Species}, \emph{dbo:Work}, \emph{foaf:Person}. The above entity types are the most common according to the total number of entities present in all 11 releases. Table~\ref{3DBpediaEntity} presents the breakdown of entity count per class. We also explored the Spanish version of DBpedia to further validate our completeness measure.
In Table~\ref{tab:dboPlace}, we present the \textit{dbo:place} class entity count across the seven releases of the Spanish DBpedia.

\begin{table*}[!ht]
	\footnotesize
	\centering
	\caption{English DBpedia 10 Classes entity count.}
	\label{3DBpediaEntity}
	\footnotesize
	\begin{tabular}{@{}rcccccccccc@{}}
		\toprule
		Version & dbo:Animal & dbo:Artist & dbo:Athlete & dbo:Film & dbo:MusicalWork & dbo:Organization & dbo:Place & dbo:Species & dbo:Work & foaf:Person    \\ \midrule
		3.3 & 51,809& 65,109 & 95,964 & 40,310 & 113,329 & 113,329 & 31,8017 & 11,8042 & 213,231 & 29,498\\
		3.4 & 87,543 & 71,789 & 113,389 & 44,706 & 120,068 & 120,068 & 337,551 & 130,466 & 229,152 & 30,860\\
		3.5 & 96,534 & 73,721 & 73,721 & 49,182 & 131,040 & 131,040 & 413,423 & 146,082 & 320,054 & 48,692\\
		3.6 & 116,528 & 83,847 & 133,156 & 53,619 & 138,921 & 138,921 & 413,423 & 168,575 & 355,100 & 296,595\\
		3.7 & 129,027 & 57,772 & 150,978 & 60,194 & 138,921 & 110,515 & 525,786 & 182,848 & 262,662 & 825,566\\
		3.8 & 145,909 & 61,073 & 185,126 & 71,715 & 159,071 & 159,071 & 512,728 & 202,848 & 333,270 & 1,266,984\\
		3.9 & 178,289 & 93,532 & 313,730 & 77,794 & 198,516 & 178,516 & 754,415 & 202,339 & 409,594 & 1,555,597\\
		2014 & 195,176 & 96,300 & 336,091 & 87,285 & 193,205 & 193,205 & 816,837 & 239,194 & 425,044 & 1,650,315\\ 
		201504 & 214,106 & 175,881 & 335,978 & 171,272 & 163,958 & 163,958 & 943,799 & 285,320 & 588,205 & 2,137,101\\
		\addlinespace
		201510 & 232,019 & 184,371 & 434,609 & 177,989 & 213,785 & 213,785 & 1,122,785 & 305,378 & 683,923 & 1,840,598\\	\addlinespace
		201604 & 227,963 & 145,879 & 371,804 & 146,449 & 203,392 & 203,392 & 925,383 & 301,715 & 571,847 & 2,703,493\\
		\bottomrule
	\end{tabular}
\end{table*}

\begin{table}[!hptb]
\centering
\small
\caption{Spanish DBpedia KB \textit{dbo:place} class entity count.}
\label{tab:dboPlace}
\begin{tabular}{cc}
\toprule 
\textbf{Release} & \textbf{Entity Count} \\ 
\midrule
3.8 & 321,166 \\ \addlinespace
3.9 & 345,566 \\ \addlinespace
2014 & 365,479 \\ \addlinespace
201504 & 389,240 \\ \addlinespace
201510 & 408,163 \\ \addlinespace
201604 & 659,481 \\ \addlinespace
201610 & 365,479 \\ \addlinespace
\bottomrule
\end{tabular}
\end{table}

\subsection{Completeness Evaluation}  \label{sec:expCompleteness}

In this section, we report the completeness evaluation results for both KBs. The general goal of this experimental analysis is to verify that \textit{dynamic features using periodic profiling can help to identify completeness issues}. We perform the quantitative and qualitative completeness analysis. Table \ref{tab:compCriteria} reports the criteria used  for the completeness evaluation. At first, we perform quantitative evaluation using the evolution-based completeness analysis (Section~\ref{sec:completeness}). Then, in the qualitative evaluation, we explore the causes of quality issues based on manual validation using the results from the quantitative analysis.

\begin{table}[!hptb]
\centering
\small
\caption{Criteria for completeness evaluation.}
\label{tab:compCriteria}
\begin{tabular}{lcp{5cm}}
\toprule
\textbf{Criteria} & \textbf{Value} & \textbf{Interpretation}\\ 
\midrule
Complete  &  1  &  The value of 1 implies no completeness issue present in the property.  \\ \addlinespace
Incomplete  &  0  & The value of 0 indicates completeness issues found in the property.
\\ \addlinespace
\bottomrule
\end{tabular}
\end{table}

\subsubsection{3cixty Nice}

\begin{figure}[!ht]
	\centering
	\includegraphics[width=0.8\linewidth]{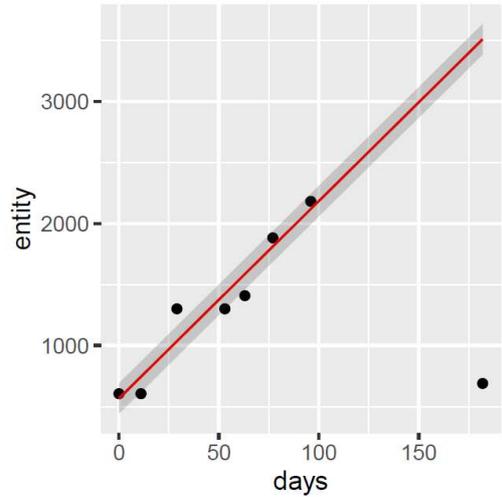}
	\caption{\textit{lode:Event} class regression line using entity counts over 8 releases.}
	\label{fig:event}
\end{figure}

\begin{figure}[!ht]
	\centering
	\includegraphics[width=0.8\linewidth]{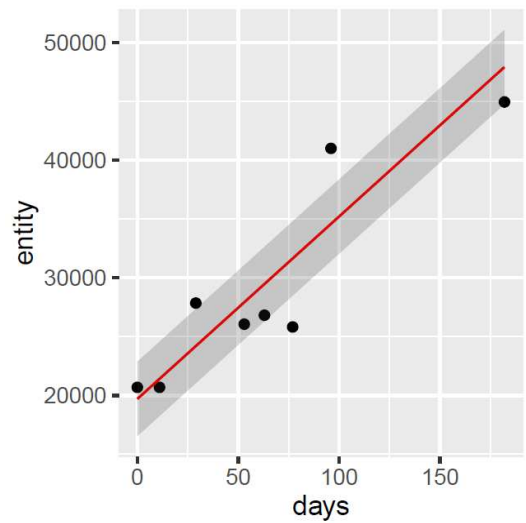}
	\caption{\textit{dul:Places} class regression line using entity counts over 8 releases.}
	\label{fig:places}
\end{figure}

Based on the entity counts reported in Table~\ref{tab:EntityCount}, we applied the linear regression over the eight releases for the \emph{lode:Event}-type and \emph{dul:Place}-type entities. We present the regression line in Figure~\ref{fig:event} and~\ref{fig:places}.

From the linear regression, the 3cixty Nice has a total of $n=8$ releases where the $8^{th}$ predicted value for \emph{lode:Event} $y^{'}_{event_{8}}=3511.548$ while the actual value=$689$. Similarly, for \emph{dul:Place} $y^{'}_{place_{8}}=47941.57$ and the actual value=$44968$.

The residuals, $e_{events_{8}}$= $|689-3511.548|=2822.545$ and $e_{places_{8}}$= $|44968-49741.57|=2973.566$.
The mean of the residuals, $e_{event_{i}}=125.1784$ and $e_{place_{i}}=3159.551$, where $i=1...n$.

So the normalized distance for the $8^{th}$ \emph{lode:Event} entity $ND_{event}$  = $\frac{2822.545}{125.1784}=22.54818$ and \emph{dul:Place} entity $ND_{place}$  = $\frac{2973.566}{3159.551}=0.9411357$.

For the \emph{lode:Event} class, $ND_{events} \geq 1$ so the KB growth measure value = $1$. However, for the \emph{dul:Place} class, $ND_{places}<1$ so the KB growth measure value =$0$ .

In the case of the 3cixty Nice KB, the \emph{lode:Event} class clearly presents anomalies as the number of distinct entities drops significantly on the last release.  
In Figure~\ref{fig:event}, the \emph{lode:Event} class growth remains constant until it has errors in the last release. It has higher distance between actual and predicted value based on the \emph{lode:Event}-type entity count. However, in the case of \textit{dul:Place}-type, the actual entity count in the last release is near to the predicted value. We can assume that on the last release the 3cixty Nice KB has improved the quality of data generation matching the expected growth.

\begin{figure*}[!ht]
	\centering
 	\includegraphics[width=1\linewidth]{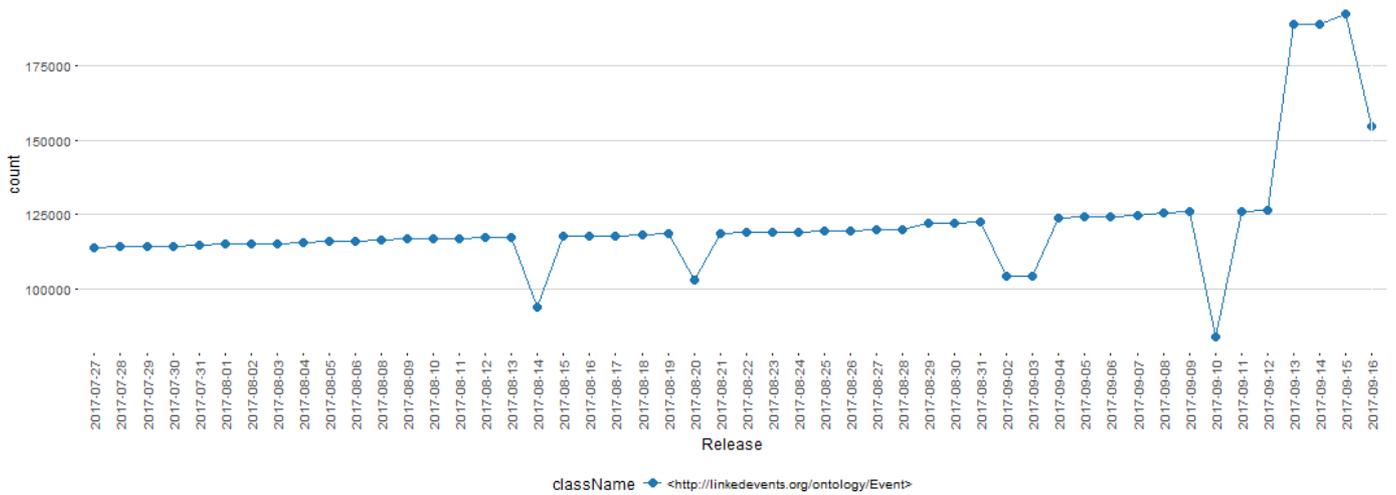}
	\caption{3cixty KB lode:Event class entity count of 50 snapshots.}
	\label{fig:3cixtyEvent}
\end{figure*}

\begin{figure*}[!ht]
	\centering
 	\includegraphics[width=1\linewidth]{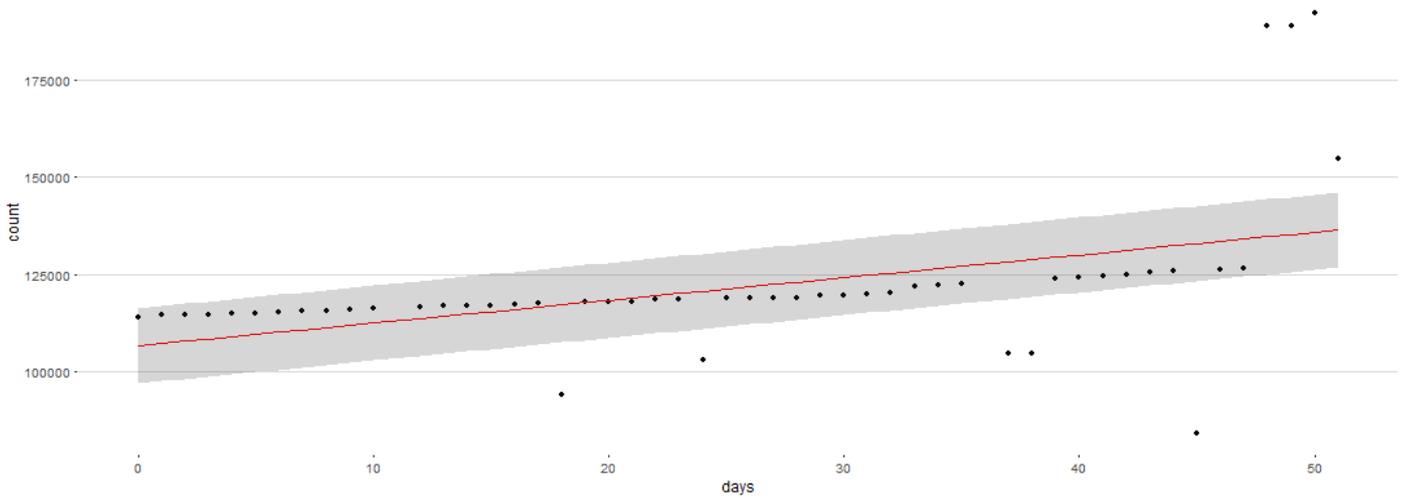}
	\caption{3cixty KB lode:Event class regression line using entity count of 50 snapshots.}
	\label{fig:3cixtyEventKBgrowth}
\end{figure*}

We then performed an empirical analysis by monitoring the 3cixty KB \textit{lode:Event} entity type. To monitor any changes present for continuous updates, we collected 50 snapshots of \textit{lode:Event} entity type from 2017-07-27 to 2017-09-16.
Table~\ref{tab:lodeEvent} reports the entity count of \textit{lode:Event} class 50 snapshots which is collected using the 3cixty KB SPARQL endpoint. Figure~\ref{fig:3cixtyEvent} illustrates the changes presents in the \textit{lode:Event}-type due to KB growth, and Figure~\ref{fig:3cixtyEventKBgrowth} reports the regression line using entity count. 
There are significant changes present in the last four releases (2017-09-13, 2017-09-14, 2017-09-15, 2017-09-16) entity count. In the 2017-09-13 release, we can see an exponential growth of actual entity count value of 190,187 compared to predicted value of 125,100 . Furthermore, on the next two releases (2017-09-14,2017-09-15) entity count remains stable due to fewer variation presents in the entity count. However, on the 2017-09-16 snapshots, we can observe a drop in the entity count which may lead to anomalies in the data integration pipeline. We further investigated the value chain leading to the generation of the KB, and we found an error in the external data acquisition process that led to missing entities for the 2017-09-16 snapshot.

To validate our assumptions, we perform property completeness measure based on the last two KB releases, namely 2016-05-15 and 2016-09-09. In Table \ref{tab:lodeEventC}, we present a subset of completeness measure results. For the \emph{lode:Event} entity type, the number of predicates in the last two releases $= 21$ and the number of predicates with completeness issues (value of 0) $= 8$. For instance, \textit{lode:Event} class property \textit{atPlace} has a frequency of $(1,632,424)$ for the releases 2016-05-15 and 2016-09-09. Based on the condition of completeness measure (Table~\ref{tab:compCriteria}), the property \textit{lode:atPlace} indicates a completeness issue.
In Table~\ref{tab:lodeEvent2}, we present completeness measures based on 50 periodic snapshots. 
For example, based on the frequency count of \textit{lode:BusinessType} in the 2017-09-15 snapshot the observed value is $(1,74,421)$ lower than 2017-09-16 snapshots value $(99,996)$. In this account, the completeness measure value is 0 leading to possible quality issues.

\begin{table}[!ht]
\centering
\small
\caption{Completeness measure of the 3cixty Nice \textit{lode:Event} class.}
\label{tab:lodeEventC}
\begin{tabular}{lccc}
\toprule
\textbf{Property} &  \textbf{2016-05-15} & \textbf{2016-09-09} & \textbf{Complete}\\ 
\midrule
lode:atPlace & 1,632 & 424 &0\\ \addlinespace
lode:atTime & 2,014 & 490 & 0\\ \addlinespace
lode:businessType & 2,182 & 689 &0\\ \addlinespace
lode:hasCategory & 1,698 & 584 & 1\\ \addlinespace
\midrule
\multicolumn{4}{c}{other rows are omitted for brevity} \\ \addlinespace
\midrule
lode:involvedAgent & 266 &  42 & 0\\ \addlinespace
\bottomrule
\end{tabular}
\end{table}

\begin{table*}[!ht]
\centering
\small
\caption{Completeness measure of 3cixty Nice \textit{lode:Event} class properties from periodic snapshots.}
\label{tab:lodeEvent2}
\begin{tabular}{lccc}
\toprule
\textbf{Property} &  \textbf{2017-09-15} & \textbf{2017-09-16} & \textbf{Complete}\\ 
\midrule
lode:minDistanceNearestWeatherStation & 2,067 & 2,063 &0\\ \addlinespace
lode:nearestWeatherStation & 2067 & 2063 & 0\\ \addlinespace
lode:businessType & 1,74,421 & 99,996 &0\\ \addlinespace
lode:minDistanceNearestMetroStation & 72,606 & 72,606 & 1\\ 
\midrule
\multicolumn{4}{c}{other rows are omitted for brevity} \\ 
\midrule
lode:created & 118,070 & 43,861 & 0\\ \addlinespace
\bottomrule
\end{tabular}
\end{table*}

\subsubsection{DBpedia}

\begin{figure}[!ht]
	\centering
	\includegraphics[width=1\linewidth]{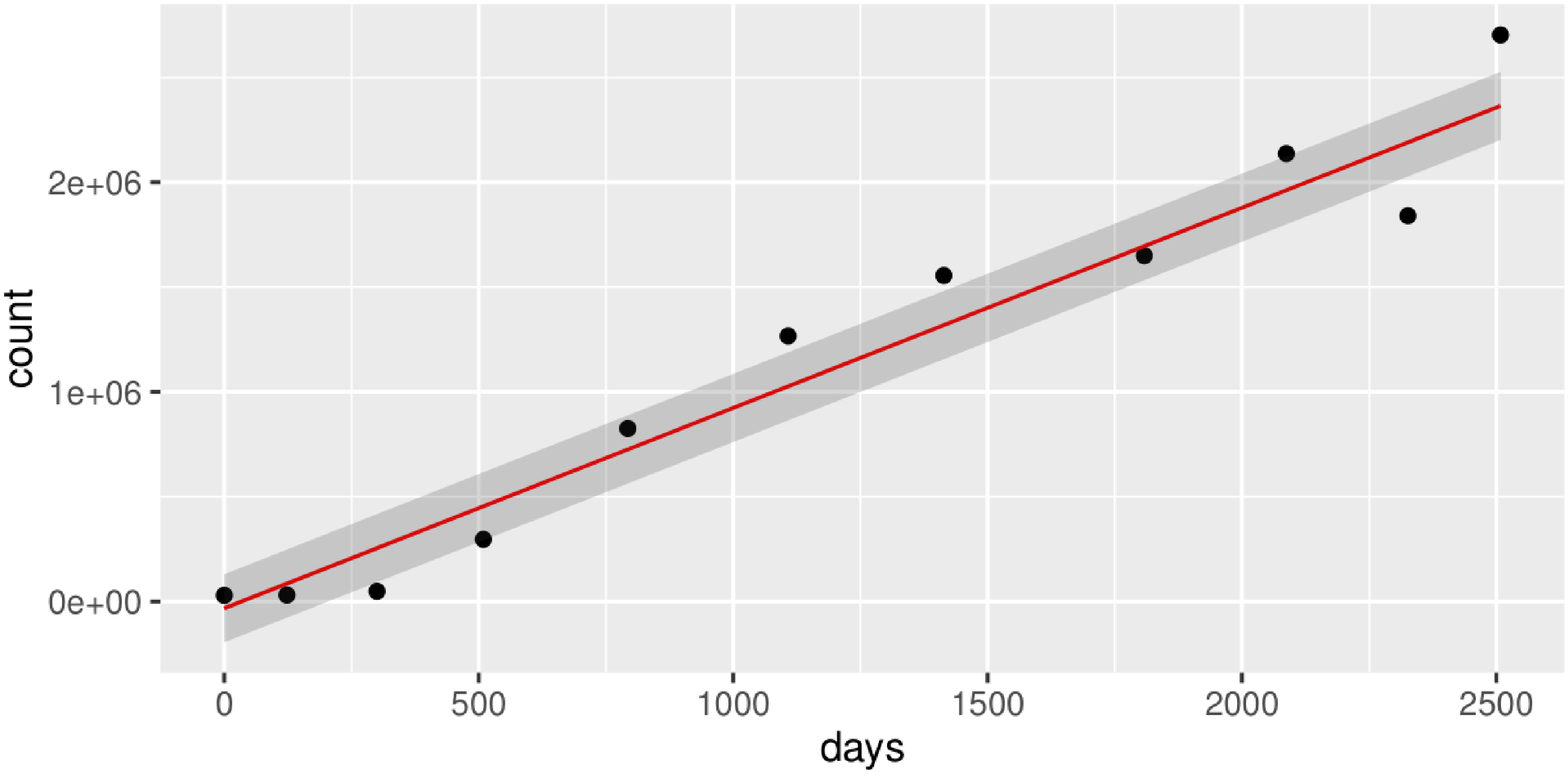}
	\caption{\textit{foaf:Person} class regression line using entity counts over 11 releases.}
	\label{fig:StabilityDBpedia}
\end{figure}

We evaluate the KB update trends based on linear regression analysis by comparing with actual and predicted values. In this account, we measured the normalized distance (ND) for each class. Based on the normalized distance, we then classify the growth of the class.
Based on the entity counts reported in Table~\ref{3DBpediaEntity}, we applied the linear regression for each class. Table~\ref{DBpedia_stability} illustrates the normalized distance and predicted growth values for each class. 
From the results observed for \textit{dbo:Artist}, \textit{dbo:Film}, and \textit{dbo:MusicalWork}, the normalized distance is near the regression line with $ND<1$. We assume that these classes have stable growth. On the contrary, \textit{dbo:Animal}, \textit{dbo:Athelete}, \textit{dbo:Organisation}, \textit{dbo:Place},\textit{dbo:Species},\textit{dbo:Work}, and \textit{foaf:Person}, the normalized distance is far from the regression line with $ND>1$. We assume that on the last release these classes might have unstable growth which may lead to completeness issue. For example, Figure~\ref{fig:StabilityDBpedia} reports the \textit{foaf:Person} class regression line using entity counts over 11 releases. 
The \emph{foaf:Person}-type last release (201604) entity count has a higher growth (over the expected). In particular, \emph{foaf:Person} has KB growth measure of 1 with a normalized distance $ND=2.08$. From this measure, we can perceive that in \emph{foaf:Person} there is completeness issue. We can imply that additions in a KB can also be an issue. It can be due to unwanted subjects or predicates.

\begin{table}[!ht]
	\centering
	\small
	\tabcolsep=0.11cm
	\caption{DBpedia 10 class Summary.} 
	\label{DBpedia_stability}
	\begin{tabular}{lcc}

		\toprule
		
		\textbf{Class} & \textbf{Normalized Distance(ND)} & \textbf{Growth}\\ \midrule
		
		dbo:Animal & 3.05 & 1    \\
		
		dbo:Artist  & 0.66 & 0 \\
		
		dbo:Athlete  & 2.03 & 1  \\
		
		dbo:Film & 0.91 & 0 \\
		
		dbo:MucsicalWork & 0.56 & 0  \\
		
		dbo:Organisation & 2.02 & 1 \\
		
		dbo:Place  & 5.03 &  1  \\
		
		dbo:Species & 5.87 & 1   \\
		
		dbo:Work  & 1.05 & 1 \\
		
		foaf:Person  & 2.08 & 1 \\
		
		\bottomrule  \\ 
		
	\end{tabular}

\end{table}


To further evaluate our assumption, we perform property completeness analysis based on the last two DBpedia KB releases of 201510 and 201604.
Table~\ref{tab:dboFoafPerson} reports the results of completeness based on the latest two releases of DBpedia 201510 and 201604 for \textit{foaf:Person} entity type. \emph{foaf:Person} has a total of $436$ properties over the two considered versions. The number of consistent properties is $238$. Based on the completeness criteria (Table ~\ref{tab:compCriteria}), we computed the completeness measures over those $238$ properties and identified $50$ properties with completeness measure value of 0. The remaining 188 properties can be considered as complete. For example, the \emph{foaf:Person} class property \textit{dbo:firstRace} has an observed frequency of 796 in the 201510 release, while it is 788 in the 201604 release. As a consequence the Completeness measure evaluated to 0; thus it indicates an issue of completeness in the KB. We further validated our results through manual validation.
Table~\ref{DBpedia_predicate_completeness} reports, for each class, the total number of properties -- which were detected by completeness computation --, the complete properties, the incomplete properties and percentage of complete properties.

Taking into account the Spanish DBpedia on the last two releases (201604, 201610) there are in total $8,659$ common properties present in the datasets. We identified $3,606$ properties with quality issues based on the frequency difference between two releases. In Table~\ref{tab:placeDBO}, we present a subset of completeness measure results. We have detected quality issues based on the property frequency difference between two versions of the \textit{dbo:Place} class. For example, the property \textit{dbo:anthem} count is $316$ for the 201610 release while it was $557$ in the 201604 release. This variation in the property count implies that $241$ resources are missing in the 201610 version of the DBpedia 201610 release. We further validated this result through manual validation.

\begin{table}[!hptb]
\centering
\small
\caption{Spanish DBpedia \textit{dbo:Place} class completeness measure based on release 201604 and 201610.}
\label{tab:placeDBO}
\begin{tabular}{lccc}
\toprule
\textbf{Property} & \textbf{201604}& \textbf{201610} & \textbf{Complete} \\ 
\midrule
dbo:abstract & 363,572  & 655,233 & 1 \\ \addlinespace
dbo:address & 17,636 & 13,3781  & 0\\ \addlinespace
dbo:anthem & 557 & 316 & 0\\ \addlinespace
dbo:archipelago & 3,162 & 1,871 & 0\\ \addlinespace
dbo:architect & 4,580 & 2,291 & 0\\ \addlinespace
dbo:architecturalStyle & 6,919 & 4,373 & 0\\ 
\midrule
\multicolumn{4}{c}{other rows are omitted for brevity} \\ 
\midrule
dbo:area & 6,764 & 3,619 & 0\\ \addlinespace
\bottomrule
\end{tabular}
\end{table}

\begin{table}[!ht]
\centering
\small
\caption{Completeness measure of DBpedia KB \textit{foaf:Person} class.}
\label{tab:dboFoafPerson}
\begin{tabular}{lccc}
\toprule
\textbf{Property} & \textbf{201510} & \textbf{201604} & \textbf{Complete}\\ 
\midrule
dbo:timeInSPace & 465 & 419 & 0\\ \addlinespace
dbo:height & 139,445 & 148,192 & 1\\ \addlinespace
dbo:weight & 67,412 & 66,144 &0\\ \addlinespace
dbo:abstract & 1,282,025 & 1,165,251 & 0\\ 
\midrule
\multicolumn{4}{c}{other rows are omitted for brevity} \\ 
\midrule
dbo:activeYearsEndDate & 26,483 &  25,221 & 0\\ \addlinespace
dbo:firstRace & 796 &  788 & 0\\ \addlinespace
\bottomrule
\end{tabular}
\end{table}

\begin{table}[!ht]
	\centering
	\small
	\tabcolsep=0.11cm
	\caption{DBpedia 10 class completeness measure results based on release 201510 and 201604.} 
	\label{DBpedia_predicate_completeness}
	\begin{tabular}	
				{p{1.5cm}rrrrr}
		\toprule

		\textbf{Class} & \textbf{Properties} & \textbf{Incomplete} & \textbf{Complete} & \textbf{Complete(\%)}\\ \midrule
		
		dbo:Animal & 170  & 50 & 120 & 70.58\% \\
		
		\addlinespace
		
		dbo:Artist & 372   & 21 & 351 & 94.35\% \\
		\addlinespace
		
		dbo:Athlete & 404  & 64 & 340 & 84.16\%\\
		\addlinespace
		
		dbo:Film & 461 & 34 & 427 & 92.62\% \\
		\addlinespace
		
		dbo:MusicalWork & 335  & 46 & 289 & 86.17\% \\
		\addlinespace
		
		dbo:Organisation & 975 & 134 & 841 & 86.26\%\\
		\addlinespace
		
		dbo:Place &  1,060  & 141 & 920 & 86.69\% \\
		\addlinespace
		
		dbo:Species & 101  & 27 & 74 & 73.27\%\\
		\addlinespace
		
		dbo:Work & 896  & 89 & 807 & 90.06\%\\
		\addlinespace
		
		foaf:Person & 396 & 131 & 265 & 66.92\%\\
		
		\bottomrule  \\ 
		
	\end{tabular}

\end{table}

\subsection{Manual Validation} 

We manually inspected whether the detected issues by the Feature Extraction stage are real issues. We annotated each property either as True positive (TP) or False positive (FP) (Sec.~\ref{sec:ManualValidation}). 

Taking into consideration the results from completeness analysis, we randomly selected a subset of properties to make the task feasible to be performed manually.\footnote{We remind here that the intent is to be precise, rather than maximizing the quantity of the annotations. We have studied empirically that a good number of annotations is 250 and demonstrated by the experimental results.} 
Concerning the 3cixty Nice KB, we analyzed the properties attached to \emph{lode:Event} entities. On the other hand for the English and Spanish DBpedia KB, we explored \textit{dbo:Place} and \textit{foaf:Person} entity types. The completeness manual validation results are explained in detail below.
 
 
\textbf{\emph{lode:Event} properties:} 
In the last two releases of \textit{lode:Event} class we found 21 common properties. From this list, we found only eight properties have completeness value of 0. 

\textit{Instances.} We investigated all entities attached to this eight properties and we extracted five instances for each property, in total we manually collected 40 different entities.

\textit{Inspection.} We observed that entities that are present in 2016-06-06 are missing in 2016-09-09. Thus, it leads to a completeness value of 0. 
As a result we identified a total of $1,911$ entities missing in the newest release: this is an actual error. We further investigated and found an error in the reconciliation algorithm for 2016-09-09 release.
In this account, the variation present in the stability measures is true positive.
Furthermore, based on the True Positive and False Positive results, the output from completeness measure has a precision of $95\%$.

\textbf{\emph{foaf:Person}/\emph{dbo:firstRace} property:} 
For the \textit{foaf:Person} entity type, we found 238 common properties in last two releases (201510, 201604) for the English DBpedia KB.
From the completeness measure over 396 properties only $131$ properties have a completeness value of 0.

\textit{Instances.} We investigated a subset of $50$ incomplete properties based on the subjects present for each property. For example, property \textit{dbo:firstRace} and \textit{dbo:lastRace} have completeness value of 0. We extracted all the subjects present in the last two releases (201510 and 201604) and performed a set disjoint operation to identify the missing subjects. For manual validation, we first checked five subjects for the \textit{dbo:firstRace} and \textit{dbo:lastRace} property, checking a total of $250$ entities.

\textit{Inspection.} In the 201604 release, \textit{dbo:firstRace} has $769$ instances and in the 201510 release it has $777$ instances. After the set disjoint operation between two releases (201510, 201604), we found $9$ distinct instances missing in 201604 release of the English DBpedia version.
Furthermore, we manually inspected each instance to identify causes of incompleteness issue. One of the data instance \textit{dbr:Bob\_Said} for the \textit{dbo:firstRace} property is available in the 201510 release. However, it is not present in 201604 release. We further explore the corresponding Wikipedia page using \textit{foaf:primaryTopic}. In the Wikipedia page \textit{firt race} is present as info box key. Due to DBpedia update from 201510 to 201604 version, this entity has been missing from the property \textit{dbo:firstRace}. Similarly, we also found this entity is missing for the \textit{dbo:lastRace} property.
This presents an ideal scenario for completeness issues in the 201604 release of the English version of DBpedia. Based on the manual inspection of 50 properties, we observed that completeness measure has the precision of $94\%$.

\textbf{\emph{dbo:Place/\textit{dbo:prefijoTelef\'{o}nicoNombre}} property:}
From the Spanish version of DBpedia, \textit{dbo:Place} entity type completeness measure we found $3,606$ properties with completeness value of 0. This indicates a potential completeness issue present for these properties.

\textit{Instances.} From the $3,606$ property, we randomly selected the property \textit{dbo:prefijoTelef\'{o}nicoNombre} for manual validation. We collected all the subjects (56109, 55387) from the two releases (201604, 201610). Then we performed a set of disjoint operations between two triples set to identify those triples missing from the 201610.

\textit{Inspection.} From the set disjoint operation, we found a total of 1982 subject missing from 201610 version. To keep the manual work at a feasible level, we selected a subset of 200 subjects for evaluation in a random manner. One of the results of the analysis is location \textit{Morante},\footnote{\url{http://es.dbpedia.org/page/Morante}} which is available in the 201604 release. However, it is missing in 201610 release of DBpedia. To further validate such an output, we checked the source Wikipedia page using \textit{foaf:primaryTopic} about \textit{Morante}.\footnote{\url{https://es.wikipedia.org/wiki/Morante}}
In the Wikipedia page \textit{prefijo Telef\'{o}nicoNombre} is present in the infobox as key. In the Spanish DBpedia from 201604 version to 201610 version update, this subject has been missing from the property \textit{prefijo Telef\'{o}nicoNombre}. This example shows a completeness issue presents in the 201610 release of DBpedia for property \textit{prefijo Telef\'{o}nicoNombre}.  
Based on the investigation over the subset of property values, we compute our completeness measure has the precision of $89\%$.

\subsection{Consistency Evaluation} \label{sec:expValidate}

The process leading to the constraint definitions is outlined in Section~\ref{sec:integrityConstraints}. 
From the quantitative analysis, we have identified multiple entity types and properties with quality issues. In particular, we selected the entity types from the completeness analysis for consistency analysis and evaluated the performance of the constraint classifier using five learning models. Our approach has been implemented with a prototype written in R.\footnote{\label{note:ShapeInduction}~\url{https://github.com/rifat963/RDFShapeInduction}}

\textbf{\textit{Feature Extraction.}} 
We are evaluating the integrity constraints evaluation as a classification problem, it is necessary to further validate the annotations and create a gold standard. In this context, we have manually inspected the constraints feature (Section \ref{FeatureEngineering}) values from the 3cixty and DBpedia KB. However, to keep the manual inspection tasks at the feasible level, we have selected a subset of properties for an entity type. 

In this experimental analysis for the English DBpedia KB, we used the expected cardinalities for 174 properties (associated with an instance of a given class). Also, we collected a subset of 200 properties associated with the \textit{dbo:Place} entity type for IRI objects and the datatype for literal objects. Similarly, for Spanish DBpedia we collected cardinality features for 240 properties and 219 properties for the range constraints based on \textit{dbo:Organization} entity type. 
Furthermore, we collected dataset with cardinality features for each property associated with instances of a given class for 215 properties for the 3cixty Nice KB. For range constraints, we collected 215 properties associated with IRI and the datatype for literal objects.
Following we present an example of feature extraction process based on minimum cardinality, maximum cardinality and range constraints.

\begin{description}


  \item[Cardinality constraints:] 
   We generate cardinality information for each property associate with the instances of a given class.  For example, by analyzing 1,767,272 \textit{dbo:Person} instances in DBpedia, we extract the cardinality distribution for \textit{dbo:Person-dbo:deathDate} as reported in Table~\ref{tab:rowCard}.
    
    \begin{table}[!hptb]
\centering
\small
\caption{Cardinality Counts for \textit{dbo:Person-dbo:deathDate}.}
\label{tab:rowCard}
\begin{tabular}{lll}
\toprule
\textbf{Cardinality} & \textbf{Instances} & \textbf{Precentage} \\ 
\midrule
0 & 1,355,038 & 76.67\% \\ \addlinespace
1 & 404,069 & 22.87\% \\ \addlinespace
2 & 8,165 & 0.46\% \\ \addlinespace
\bottomrule
\end{tabular}
\end{table}

During the feature extraction step, this raw profiling data is used to derive a set of features that can be used for predicting the cardinality. Another example of cardinality distribution is reported in Table~\ref{tab:rowCardUnion} for the \textit{dbo:Sport/dbo:union} property.

\begin{table}[!hptb]
\centering
\small
\caption{Cardinality Counts for \textit{dbo:Sport/dbo:union}.}
\label{tab:rowCardUnion}
\begin{tabular}{lll}
\toprule
\textbf{Cardinality} & \textbf{Instances} & \textbf{Precentage} \\ 
\midrule
0 & 1,662 & 84.88\% \\ \addlinespace
1 &   279 & 14.14\% \\ \addlinespace
2 &    10 & 0.05\% \\ \addlinespace
3 &     5 & 0.02\% \\\addlinespace
4 &     2 & 0.01\% \\ \addlinespace
\bottomrule
\end{tabular}
\end{table}

At first we extract the raw cardinalities. Based on the raw values, we compute the distinct cardinality values distributions similar to the ones reported in Table~\ref{tab:rowCardUnion}. Note that there are three distributions, one is the raw cardinalities (0,1,0,3,1,2,1,6,1,0), then distinct cardinalities (0,1,2,3,4) and the percentages of instances per each cardinality (84.88\%, 14.24\%, 0.05\%, 0.02\%, 0.01\%). 
Further, for each of the three distributions we derive 30 statistical measures including min-max cardinalities, mean, mode, standard deviation, variance, quadratic mean, skewness, percentiles, and kurtosis~\cite{freedman2009statistical}.

Table~\ref{tab:derivedFeature} reports 30 features (P1 to P30) selected for a classifier that predicts the cardinality category with example values for the \textit{dbo:Sport} class \textit{dbo:union} property. Features P1 to P13 are related to raw cardinality distribution, features P14 to P20 are related to the distinct cardinality distribution, and features P21 to P30 are related to the percentage distribution. For example, P1 presents a minimum cardinality value of 0 for \textit{dbo:Sport/dbo:union} and P2 presents maximum that is 4. Our intuition is that these are descriptive to classify the cardinality category. Nevertheless, the data can be noisy and either min or/and max could be outliers. To address this we add statistical features that give more insights about the distribution of the cardinalities such as mean, mode, kurtosis, standard deviationsm, skewness, variance and four percentiles. Our motivation for using these statistical values is that each of these could provide some insights related to different possible cardinality distributions. Based on the cardinality level (Sec.~\ref{sec:integrityConstraints}), we create a gold standard by annotating the properties with corresponding constraints values and create the feature dataset for validation.
For instance, the \textit{dbo:Person-dbo:deathDate} corresponding SHACL property constraints are generated as illustrated by Listing~\ref{lst:cardcons}.

\begin{table}[!ht]
\centering
\footnotesize
\caption{\textit{dbo:Sport/dbo:union} 30 statistical measures (p1 to p30) from raw cardinality estimation.}
\label{tab:derivedFeature}
\begin{tabular}{lp{1.5cm}clp{1.5cm}c}
\toprule
\textbf{ID} & \textbf{Description} & \textbf{Example} & \textbf{ID} & \textbf{Description} & \textbf{Example}  \\ 
\midrule
P1 & Min Cardinality & 0 & P16 & Distinct Quadratic Mean & 2.4495 \\ \addlinespace
P2 & Max Cardinality & 4 & P17 & Distinct Kurtosis & -1.2 \\ \addlinespace
P3 & Mean & 0.16445 & P18 & Distinct Standard Deviation & 1.5811 \\ \addlinespace
P4 & Mode & 0 & P19 & Distinct Skewness & 0 \\ \addlinespace
P5 & Quadratic mean & 0.44972 & P20 & Distinct variance & 2.5 \\ \addlinespace
P6 & Kurtosis & 13.7897 & P21 & Percentages Mins & 0.0010 \\ \addlinespace
P7 & Standard Deviation & 0.41868 & P22 & Percentage Max & 0.8488 \\ \addlinespace
P8 & Skewness & 3.09484 & P23 & 0 Percentage & 0.8488 \\ \addlinespace
P9 & Variance & 0.17529 & P24 & 1 Percentage & 0.1429 \\ \addlinespace
P10 & 98th percentile & 1 & P25 & Percentage Mean & 0.2 \\ \addlinespace
P11 & 2nd percentile & 0 & P26 & Percentage Quad. Mean & 0.3849 \\ \addlinespace
P12 & 75nd percentile & 0 & P27 & Percentage Kurtosis & 0.3849 \\ \addlinespace
P13 & 25th percentile & 0 & P28 & Percentage Standard Deviation & 0.3677 \\ \addlinespace
P14 & Distinct Cardinalities & 5 & P29 & Percentage Skewness & 2.0948 \\ \addlinespace
P15 & Distinct Mean Card. & 0 & P30 & Percentage Variance & 0.1352 \\ \addlinespace
\bottomrule
\end{tabular}
\end{table}

\item[Range Constraints:] We collected statistics about the number of IRIs, Literals, and Blank nodes for each property associated with instances of a given class as shown in Table~\ref{tab:rangeLoupe}. The blank node counts are also generated by the data collection stage but they are not reported because there were no blank nodes in this example.

\begin{table}[!ht]
\centering
\footnotesize
\caption{Object node type information.}
\label{tab:rangeLoupe}
\begin{tabular}{lrrrr}
\toprule
\multirow{2}{*}{\textbf{Class-property}} & \multicolumn{2}{c}{\textbf{IRI}} & \multicolumn{2}{c}{\textbf{Literals}} \\ \cline{2-5} 
 & \textbf{Total} & \textbf{Distinct} & \textbf{Total} & \textbf{Distinct} \\ 
 \midrule
{\small dbo:Person/dbp:birthPlace} & 89,355 & 21,845 & 44,639 & 20,405 \\ \addlinespace
{\small dbo:Person/dbp:name} & 21,496 & 15,746 & 115,848 & 100,931 \\ \addlinespace
{\small dbo:Person/dbp:deathDate} & 127 & 111 & 65,272 & 32,449 \\ \addlinespace
{\small dbo:Person/dbp:religion} & 8,374 & 786 & 6,977 & 407 \\ \addlinespace
\bottomrule
\end{tabular}
\end{table}

Furthermore, we also explore object type information by analyzing all IRI and blank node objects. Table 4 shows an example of object type information by analyzing all objects having  \textit{dbo:Person/dbp:deathPlace} as class-property. 
As it can be seen, the objects of \textit{dbo:Person/dbp:deathPlace} are typed as many different classes. And, in general, it can be seen that most objects are typed with multiple classes (e.g., with equivalent classes, super classes). Also there are some objects that should not be associated (\textit{i.e.}, inconsistent) with the \textit{dbp:deathPlace} property, for example, a \textit{Broadcaster} should not be a death place of a person. Further, there are some objects for which the type information is not available.  

\begin{table}[!ht]
\centering
\footnotesize
\caption{Classes of \textit{dbo:Person-dbp:birthPlace} objects.}
\label{tab:objcls}
\begin{tabular}{lllll}
\toprule
\multirow{2}{*}{\textbf{Object Class}} & \multicolumn{2}{c}{\textbf{\begin{tabular}[c]{@{}c@{}}Objects\\ (89,355)\end{tabular}}} & \multicolumn{2}{c}{\textbf{\begin{tabular}[c]{@{}c@{}}Distinct Objects\\ (21,845)\end{tabular}}} \\ \cline{2-5} 
 & Count & \% & Count & \% \\ 
\midrule
schema:Place & 71,748 & 80.29 & 16,502 & 75.54 \\ \addlinespace
dbo:Place & 71,748 & 80.29 & 16,502 & 75.54 \\ \addlinespace
dbo:PopulatedPlace & 71,542 & 80.07 & 16,353 & 74.86 \\ \addlinespace
dbo:Settlement & 41,216 & 46.13 & 14,184 & 64.93 \\ \addlinespace
\midrule
\multicolumn{5}{c}{other rows are omitted for brevity} \\ 
\midrule
\addlinespace
schema:Product & 2 & 00.00 & 2 & 00.01 \\ \addlinespace
dbo:Broadcaster & 2 & 00.00 & 2 & 00.01 \\ \addlinespace
Unknown & 9,790 & 10.95 & 2,888 & 13.22 \\ \addlinespace
\bottomrule
\end{tabular}
\end{table}

Similarly, for literal objects our data collection module extracts the information about their data types. Table~\ref{tab:object-datatypes} shows an example of extracted information for the class-property combination \textit{dbp:Person/dbp:deathDate}. For each datatype, it shows the number of objects, number of distinct objects, and their corresponding percentages. Such an information provides heuristics about which should be the corresponding datatype. 

\begin{table}[!ht]
\centering
\footnotesize
\caption{Datatypes of \textit{dbp:Person/dbp:deathDate} literals.}
\label{tab:object-datatypes}
\begin{tabular}{lllll}
\toprule
\multirow{2}{*}{\textbf{Datatype}} & \multicolumn{2}{c}{\textbf{\begin{tabular}[c]{@{}c@{}}Objects\\ (65,272)\end{tabular}}} & \multicolumn{2}{c}{\textbf{\begin{tabular}[c]{@{}c@{}}Distinct Objects\\ (32,449)\end{tabular}}} \\ \cline{2-5} 
 & Count & \% & Count & \% \\ 
\midrule
xsd:date & 39,761 & 60.92 & 26,726 & 82.36 \\ \addlinespace
xsd:integer & 13,543 & 20.75 & 1,758 & 5.42 \\ \addlinespace
rdf:langString & 6,388 & 9.79 & 3,512 & 10.82 \\ \addlinespace
xsd:gMonthDay & 5,446 & 8.34 & 366 & 1.13 \\ \addlinespace
dt:second & 113 & 0.17 & 66 & 0.20 \\ \addlinespace
xsd:double & 20 & 0.03 & 20 & 0.06 \\ \addlinespace
dt:hour & 1 & 0.00 & 1 & 0.00 \\ \addlinespace
Total & 65,272 & 100 & 32,449 & 100 \\ \addlinespace

\bottomrule
\end{tabular}
\end{table}

We use all the aforementioned information as features for the two tasks of detecting the object type and also detecting the class type for IRI objects and the datatype for literal objects. 
    
\item[String constraints:] We use statistics about the literals to identify the minLength and maxLength of the string values. 
Based on the string length distribution of literal values, we explore the 1st quartile and 3rd quartile to identify the minimum and maximum length.
More specifically, we evaluate the interquartile range (IQR) based on the string length literal values of a property.
For example, in Table~\ref{tab:stringLengthTitle}, we report the string length distribution of the \textit{foaf:Person} class \textit{dbo:Title} property together with frequency of string length. 
Similarly, in Table~\ref{tab:stringLengthName}, it is illustrated the \textit{dbo:BirthName} property frequency distribution.

\begin{table}[!hptb]
\small
\centering
\caption{Frequency distribution of \textit{foaf:Person}/\textit{dbo:Title} property.}
\label{tab:stringLengthTitle}
\begin{tabular}{lll}
\toprule
\textbf{String Length} & \textbf{Frequency} & \textbf{Percentage} \\ 
\midrule
16 & 20 & 31.25 \% \\ \addlinespace
13 & 7 & 10.93\% \\ \addlinespace
15 & 5 & 7.81\% \\ \addlinespace
\midrule
\multicolumn{3}{c}{other rows are omitted for brevity} \\ 
\midrule \addlinespace
20 & 4 & 6.25\% \\ \addlinespace
\bottomrule
\end{tabular}
\end{table}
\begin{table}[!hptb]
\small
\centering
\caption{Frequency distribution of \textit{foaf:Person}/\textit{dbo:BirthName} property.}
\label{tab:stringLengthName}
\begin{tabular}{lll}
\toprule
\textbf{String Length} & \textbf{Frequency} & \textbf{Percentage} \\ 
\midrule
20 & 32 & 13.14 \% \\ \addlinespace
21 & 26 & 10.65\% \\ \addlinespace
19 & 25 & 10.24\% \\ \addlinespace
\midrule
\multicolumn{3}{c}{other rows are omitted for brevity} \\
\midrule
\addlinespace
22 & 17 & 6.96\% \\ \addlinespace
\bottomrule
\end{tabular}
\end{table}
In this example, both properties have a small central tendency towards the mean. Our main focus is to identify a range of minLength and maxLength for literal objects. In this account, we use the interquartile range for \textit{dbo:title} to identify minLength and maxLength. We used the 3rd quartile (Q3) of the string length as maxLength and the 1st quartile (Q1) as minLength for the \textit{dbo:title} property. In Figure~\ref{fig:boxplot}, we present a boxplot of the \textit{dbo:title} property. In particular, using the interquartile range, we can present the string range constraints as a binary classifier. 

 \begin{figure}[!ht]
	\centering
	\includegraphics[width=1\linewidth]{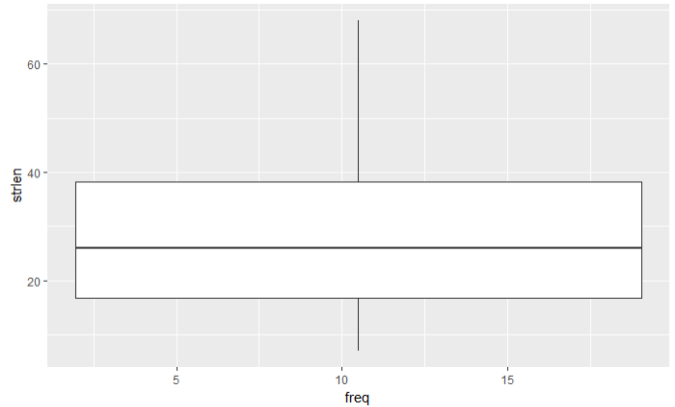}
	\caption{foaf:Person class dbo:title property string length box plot.}
	\label{fig:boxplot}
\end{figure}

\end{description}

\textbf{\textit{Model Preparation.}} 
From the initial analysis of the feature dataset, we found that the minimum cardinality constraint has an imbalance in distribution of feature values. We observed that rare events occur in case of selected constraints as response variables. The variation between two variables is less than 15\%. We applied SMOTE (Synthetic Minority Over-sampling Technique)~\cite{chawla2002smote} for oversampling the rare events. The SMOTE function over-samples response variables by using bootstrapping and k-Nearest Neighbor to synthetically create additional observations of that response variable.
In our experiment, we applied an over-sampling value of 100 to double the number of positive cases, and an undersampling value of 200 to keep half of what was created as negative cases. It balances the classifier and achieves better performance than only under-sampling the majority class. The results are reported in Table~\ref{tab:smote}.
After applying the SMOTE technique, we applied 10-fold cross-validation based on the learning models mentioned in Section \ref{sec:ML}. 

\begin{table*}[!ht]
\centering
\small
\caption{DBpedia and 3cixty Nice distribution of cardinality constraints.}
\label{tab:smote}
\begin{tabular}{cccccccc}
\toprule
\multirow{2}{*}{\textbf{Knowledge Base}}& \multirow{2}{*}{\textbf{Distribution}} & \multicolumn{2}{c}{\begin{tabular}[c]{@{}c@{}}\textbf{Minimum} \\ \textbf{Cardinality}\end{tabular}} & \multicolumn{2}{c}{\begin{tabular}[c]{@{}c@{}}\textbf{Maximum} \\ \textbf{Cardinality}\end{tabular}} & \multicolumn{2}{c}{\begin{tabular}[c]{@{}c@{}}\textbf{Range} \\ \textbf{Constraint}\end{tabular}} \\ \cline{3-8} 
                         &     & MIN0                                     & MIN1+                                     & MAX1                                     & MAX1+                                    & IRI                                 & LIT                                     \\ \midrule
 \multirow{2}{*}{3cixty Nice} & Without SMOTE                 & 47\%                                     & 52.8\%                                   & 79.2\%                                   & 20.8\%                                   & 68.7\%                                  & 31.3\%                                  \\ \addlinespace
&With SMOTE (100,200)          & 50\%                                     & 50\%                                     & 50\%                                     & 50\%                                     & 50\%                                    & 50\%                                    \\ \addlinespace 
 \multirow{2}{*}{English DBpedia} & Without SMOTE                 & 76.5\%                                   & 23.5\%                                   & 53\%                                     & 47\%                                     & 71.5\%                                  & 28.5\%                                  \\ \addlinespace
& With SMOTE(100,200)           & 50\%                                     & 50\%                                     & 50\%                                     & 50\%                                     & 50\%                                    & 50\%                                    \\\addlinespace 
\multirow{2}{*}{Spanish DBpedia}  & Without SMOTE                 & 72\%                                     & 28\%                                     & 56\%                                     & 44\%                                     & 69.4\%                                  & 30.6\%                                  \\ \addlinespace
& With SMOTE(100,200)           & 50\%                                     & 50\%                                     & 50\%                                     & 50\%                                     & 50\%                                    & 50\%                                    \\ \bottomrule
\end{tabular}
\end{table*}

\begin{table*}[!ht]
\small
\centering
\caption{Integrity Constraints performance measures for 3cixty Nice.}
\label{tab:Validation3cixty}
\begin{tabular}{ccccccclll}
\toprule
\multirow{2}{*}{\textbf{\begin{tabular}[c]{@{}c@{}}Learning\\  Algorithm\end{tabular}}} & \multicolumn{3}{c}{\textbf{Minimum Cardinality}}  & \multicolumn{3}{c}{\textbf{Maximum Cardinality}}  & \multicolumn{3}{c}{\textbf{Range}}                \\ \cline{2-10} 
\addlinespace                                                                                        & \textbf{Precision} & \textbf{Recall} & \textbf{F1} & \textbf{Precision} & \textbf{Recall} & \textbf{F1} & \textbf{Precision} & \textbf{Recall} & \textbf{F1} \\ 
\midrule
                                                                                  
\textbf{Random Forest}     &  \textbf{0.9626}     &    \textbf{0.8729}  &  \textbf{0.9156}     &   \textbf{ 0.8909}   &   \textbf{0.9423 }              &    \textbf{0.9159}         &     \textbf{0.9333}   &    \textbf{0.9032}   &\textbf{0.9180} \\ \addlinespace

Multilayer Perceptron    & 0.8812   & 0.8812    &  0.8128  &  0.8113  &  0.8269  & 0.8190  
&  0.9375  &    0.8823      &  0.9091           \\ \addlinespace

Least Squares SVM    & 0.7692  &    0.7263   &  0.7471   & 0.8070     & 0.8846     & 0.8440  &   0.8148    &     0.9167           &   0.8627           \\ \addlinespace

Naive Bayes       & 0.7152    &  0.6932      &   0.7040   &  0.7288    &     0.8268            &    0.7748         &    0.8266     &    0.7462    &  0.8275   \\ \addlinespace

K-Nearest Neighbour   & 0.6991  & 0.6695     &  0.6840     &  0.7049       &  0.8269                 &    0.7611         &      0.7837      &  0.8285            &  0.8055   \\ 
\bottomrule

\end{tabular}
\end{table*}

    \begin{table*}[!ht]
\small
\centering
\caption{Integrity Constraints performance measure for English DBpedia.}
\label{tab:ValidationENDBpedia}
\begin{tabular}{ccccccclll}
\toprule
\multirow{2}{*}{\textbf{\begin{tabular}[c]{@{}c@{}}Learning\\  Algorithm\end{tabular}}} & \multicolumn{3}{c}{\textbf{Minimum Cardinality}}  & \multicolumn{3}{c}{\textbf{Maximum Cardinality}}  & \multicolumn{3}{c}{\textbf{Range}}                \\ \cline{2-10} \addlinespace
                                                                                        & \textbf{Precision} & \textbf{Recall} & \textbf{F1} & \textbf{Precision} & \textbf{Recall} & \textbf{F1} & \textbf{Precision} & \textbf{Recall} & \textbf{F1} \\ 
                                                        \midrule                                
                                             
\textbf{Random Forest}     &  \textbf{0.9890}     &    \textbf{0.9574}  &  \textbf{0.9730}     &   \textbf{0.9842}   &   \textbf{0.9920 }              &    \textbf{0.9881}         &     \textbf{0.9457}   &    \textbf{0.9527}   &\textbf{0.9594} \\ \addlinespace

Least Squares SVM    & 0.9944  &    0.9468   &  0.9700   & 0.8491  & 0.9574 & 0.9000  &   0.8596    &     0.9231           &   0.8902           \\ \addlinespace

Multilayer Perceptron    & 0.9674   & 0.9468    &  0.9570 &  0.8167 &  0.9601  & 0.8826  
&  0.8262  &    0.8657      &  0.8456\\ \addlinespace

K-Nearest Neighbour   & 0.9511  & 0.9309     &  0.9409    &  0.8797       &  0.8750                 &    0.8773         &      0.8361      &  0.8425                 &  0.8393   \\ \addlinespace

Naive Bayes  & 0.9401    &  0.8351      &   0.8845   &  0.9065    &     0.7739            &    0.8350         &    0.8953     &    0.7951    &  0.8422   \\ 

\bottomrule
\end{tabular}
\end{table*}

\begin{table*}[]
\small
\centering
\caption{Integrity Constraints performance measure for Spanish DBpedia.}
\label{tab:ValidationDBpediaES}
\begin{tabular}{ccccccclll}
\toprule
\multirow{2}{*}{\textbf{\begin{tabular}[c]{@{}c@{}}Learning\\  Algorithm\end{tabular}}} & \multicolumn{3}{c}{\textbf{Minimum Cardinality}}  & \multicolumn{3}{c}{\textbf{Maximum Cardinality}}  & \multicolumn{3}{c}{\textbf{Range}}                \\ \cline{2-10} 
 \addlinespace                                                                                       & \textbf{Precision} & \textbf{Recall} & \textbf{F1} & \textbf{Precision} & \textbf{Recall} & \textbf{F1} & \textbf{Precision} & \textbf{Recall} & \textbf{F1} \\ 
\midrule                                                                                      
\textbf{Random Forest}     &  \textbf{0.8971}     &    \textbf{0.8547}  &  \textbf{0.8754}     &   \textbf{0.9247}   &   \textbf{0.9323}              &    \textbf{0.9285}         &     \textbf{0.8741}   &    \textbf{0.8954}   &\textbf{0.8846} \\ \addlinespace

Least Squares SVM    & 0.8517  &    0.8940   &  0.8723  & 0.8070  & 0.8846  &  0.8440  &   0.8348    &     0.8416  &  0.8381    \\ \addlinespace

Multilayer Perceptron    &  0.8670  & 0.8183  &  0.8419 &  0.8863  &  0.8517 &  0.8685 & 0.7942   & 0.7701 & 0.7819   \\ \addlinespace

K-Nearest Neighbour   & 0.8378  & 0.8170  & 0.8272  & 0.8168  & 0.7901 & 0.8032  &  0.7714  &  0.7808   & 0.7761  \\ \addlinespace

Naive Bayes   & 0.7091    & 0.7278   & 0.7183   & 0.7862  & 0.7961    &  0.7911    &  0.7620      &   0.7901   &  0.7758\\ 
\bottomrule

\end{tabular}
\end{table*}


\textbf{\textit{Model Evaluation.}} In detail, the model evaluation results are mentioned below.

\begin{itemize}
    \item \textit{3cixty Nice.} Table~\ref{tab:Validation3cixty} reports the 3cixty KB three constraints classifier performance measures. Considering five learning models, the Random Forest model had more than 90\% of F1 value for all three classifiers. For minimum cardinality, the Random Forest model reached 91\% F1 score where it achieved 96\% precision. Conversely, the Neural Network model reached 90\% F1 score for range constraints. However, simple Naive Bayes learning algorithm had a significantly lower F1 ($<$70\%) score compared to the other classifiers. K-Nearest Neighbour (K-NN) had the lowest F1 score for the maximum cardinality and range constraints.
    
    \item \textit{English DBpedia.} Table \ref{tab:Validation3cixty} illustrates the three classifiers performance measures for the English version of the DBpedia KB. Similar to the 3cixty KB, Random Forest proved to be effective in achieving greater than 90\% F1 value for all three classifiers. Overall, for Random Forest algorithm, minimum cardinality constraints reached 97\% F1 score where it achieved 98\% precision. Also, in the case of minimum cardinality classifier, other learning algorithms such as Neural Network and Least Squares SVM reached an F1 score greater than 90\%. 
    
    \item \textit{Spanish DBpedia.} Table \ref{tab:ValidationDBpediaES} reports the integrity constraints performance measure for the Spanish DBpedia Dataset. Compared to the other models, Random Forest achieved the highest F1 score for all three classifiers. In addition, it achieved 92.85\% F1 score for maximum cardinality classifier. Compared to Random Forest model, Least Squares SVM also achieved the F1 score of 87.23\%  for the minimum cardinality classifier. For the Spanish DBpedia KB, Naive Bayes classifier had the lowest F1 score for all the constraints classifiers.

\end{itemize}

\section{Discussion} \label{sec:discussion}

\begin{figure*}[!ht]
	\centering
	\includegraphics[width=1\textwidth]{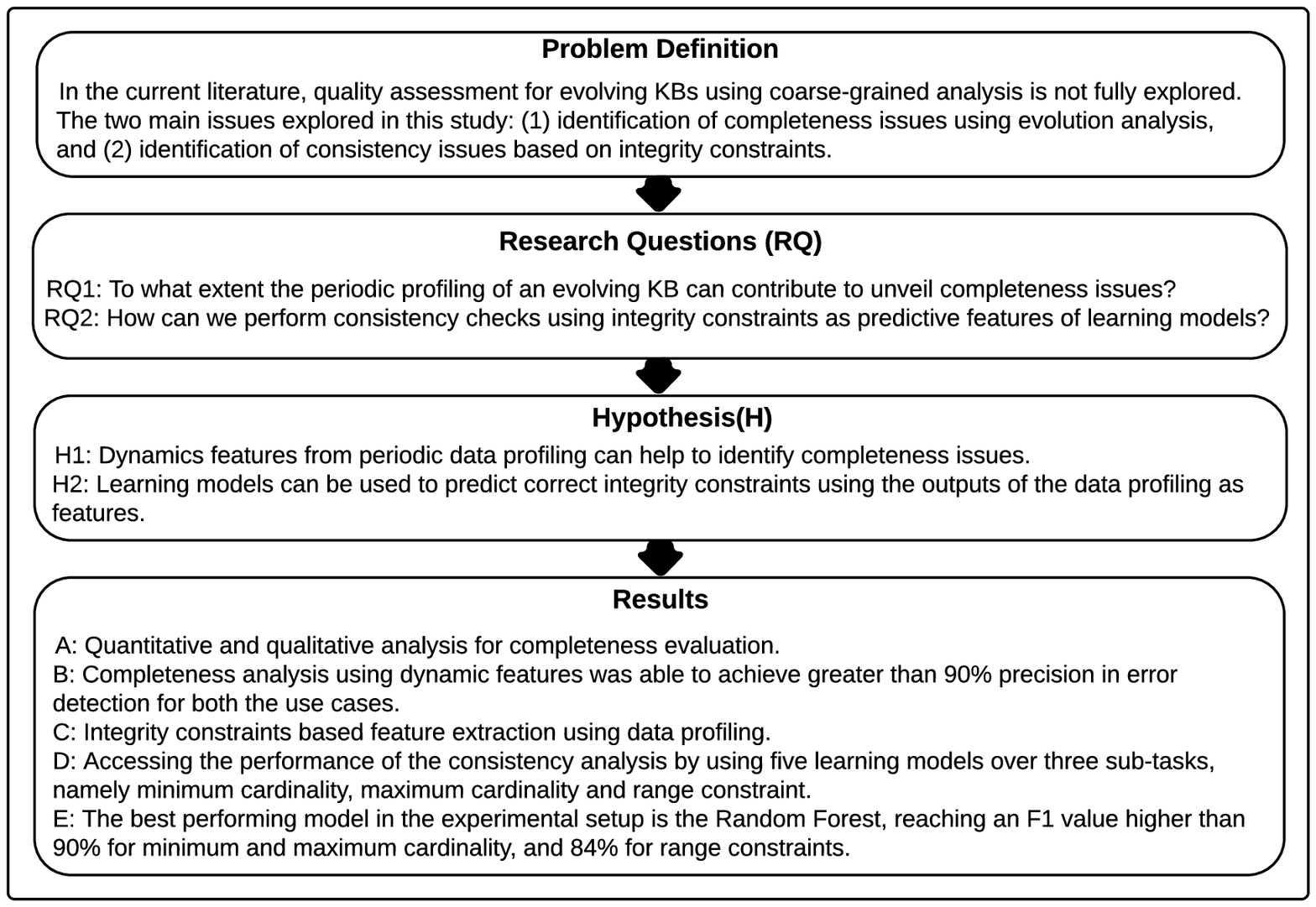}
	\caption{Summary of the main results of the Completeness and Consistency Analysis.}
	\label{fig:diusResults}
\end{figure*}

In this section, we discuss the main findings and the limitation of this work  using the results from the experimental analysis (Section~\ref{sec:experimentalAnalysis}). Figure~\ref{fig:diusResults} illustrates the primary results of this work labeled with A, B, C, D, and E. 

\subsection{Completeness Analysis}

We perceive that changes observed in a set of KB releases can help in detecting completeness issues. We identified properties with quality issues based on dynamic features from the completeness analysis. We, then, summarize our assumption using qualitative analysis by manually evaluating a subset of classes and properties. 
From the experimental analysis, we potentially detected errors in various stages of evolving KBs.  
Following we summarize our findings based on the completeness evaluation.

\begin{itemize}
    \item \textit{Causes of Quality Issues.} From our completeness evaluation, three types of quality issues are identified: \textit{(i)} errors in the data extraction process, \textit{(ii)} erroneous conceptualization, and \textit{(iii)} error in object type. In details:

    \textit{Errors in the data extraction process:}  We discovered properties with anomalies and performed further inspections for each KB. For the 3cixty KB \textit{lode:Event} entity type, we identified completeness issues due to an algorithmic error in the data extraction pipeline. For what concerns DBpedia, we identified issues as a result of missing mapping with Wikipedia infobox keys. This issue of missing mapping might happened because of wrong schema presentation or schema definition inconsistency due to KB updates.
    
    \textit{Erroneous conceptualization:} We observe that the properties with lower frequency tend to have erroneous schema representations. For example, the property \textit{dbo:weight} has 4 data instances mapped with \textit{dbo:Place} type. We further investigated each of this data instance and corresponding Wikipedia page. From manual investigation, we identified \textit{dbo:weight} property erroneously mapped with the \textit{dbo:Place} type. Such as one of the data instance \textit{wikipedia-en:Nokia\_X5} is about mobile devices, and it is mapped with \textit{dbo:Place} type. This mapping indicates a consistency issue as a result of a wrong schema presentation.

    \textit{Error in object type:} From the manual validation results, we assumed that it could be possible to identify an error in any literal value using our approach. For example, the property \textit{dbo:bnfId} triggered a completeness issue. 
    We, therefore, further investigated the property \textit{dbo:bnfId} in the 201604 release. We explored the property description that leads to Wikidata link\footnote{\url{https://www.wikidata.org/wiki/Property:P268}} and examined how \textit{BnF ID} is defined.  It is an identifier for the subject issued by BNF (Biblioth\`eque nationale de France). It is formed by eight digits followed by a check digit or letter.  Based on the \textit{BnF ID} formalization rule, we checked each literal values for \textit{dbo:bnfId} entity type. We found that one of the literal values is "12148/cb16520477z" for subject \textit{Quincy\_Davis\_(musician)}\footnote{\url{http://dbpedia.org/resource/Quincy\_Davis\_(musician)}} contains a "/" between the digits "12148" and "cb16520477z", which does not follow the standard formatting structure issued by BNF (Biblioth\`eque nationale de France). It clearly points to an error for the subject \textit{Quincy\_Davis\_(musician)}.
    However, to detect errors in literal values, we need to extend our quality assessment framework to inspect literal values computationally. We considered this extension of literal value analysis as a future research endeavor.

    \item \textit{Summary of findings.} In the case of the 3cixty Nice KB, we only identified issues based on the data source extraction process. For example, we found a significant number of resources missing for the \textit{lode:Event} class in the last release(2016-09-09). We identified all three types of quality issues for DBpedia KB. For example, entities missing in \textit{foaf:Person} class is due to incorrect mappings of field values in the data extraction process. Also, we identified a notable number of issues due to wrong schema presentation for the DBpedia KB. Such as property \textit{dbo:Lake} mapped with \textit{foaf:Person}-type due to automatic mapping with wrong Wikipedia infobox keys. Taking into account periodicity of KBs, we observe that continuously analyzing KBs with high-frequency updates (daily updates), such as the 3cixty Nice KB, has fewer quality issues. On the other hand, KBs with low-frequency updates (monthly or yearly updates), such as DBpedia KB, seem to have more completeness issues.
    
    Correspondingly, we analyze the KB growth patterns to predict any unstable behaviour. We define this lifespan analysis as \emph{stability feature}.
    A straightforward interpretation of the stability of a KB is monitoring the dynamics of knowledge base changes. This dynamic feature could be useful to understand high-level changes by analyzing KB growth patterns. However, a further exploration of the KB stability feature is needed, and we consider this as a future research activity. 
    
    Overall, we evaluated the property completeness measure in terms of precision through manual evaluation. Considering computational complexity, we only use count and difference operation for measurement functions. We assume that our computational complexity will be $O(N_{T})$ where the $N_{T}$ is the total number of entities for type T. The computed precision of completeness measure in our approach is: \textit{i)} $94\%$ for \emph{foaf:Person}-type entities of the English DBpedia KB; \textit{ii)} $89\%$ for \emph{dbo:Place}-type entities of the  Spanish DBpedia KB, and \textit{iii)} $95\%$ for the \emph{lode:Event}-type entities of the 3cixty Nice KB.

\end{itemize}

\subsection{Consistency Analysis}


In the consistency analysis, the constraint classifiers performance is measured by precision, recall and F1 score. Overall, our constraints classifiers achieved high predictive performance with the Random Forest model. For example, the Random Forest cardinality classifiers achieved the highest F1 score for all KBs. Furthermore, the Multilayer Perceptron and the Least Squares SVM also achieved high F1 scores greater than 90\% for the English DBpedia KB. Concerning the range constraints, we explored the object node type constraint for each property associated with a given class. Similar to cardinality constraints, Random Forest algorithm achieved a high F1 score of 95.94\% for the English DBpedia KB. This makes the consistency evaluation approach adaptable and facilitates adoption for multiple KBs.

Furthermore, we applied a Naive Bayes classifier. The model provides apriori probabilities of no-recurrence and recurrence events as well as conditional probability tables across all attributes.  We considered Naive Bayes as a baseline model to explore the classifier performance compared to other learning algorithms. In this context, other models achieved better performance values compared to the Naive Bayes learning algorithm.

Finally, we generate constraints once the constraint prediction models are built. 
Based on the Random Forest model, we created the constraints datasets. More specifically, we combined all the constraints related to a given class and, for each, we generate an RDF Shape. 
An example of the RDF Shape in SHACL for the \textit{foaf:Person} class is illustrated in Listing~\ref{lst:Disshape} using cardinality and range constraints. Furthermore, we perceived that the generated constraints datasets can be used in other tools such as RDFUnit~\cite{kontokostas2014test}. We considered this extension of our RDF shape induction approach as a future work.

\begin{lstlisting}[caption={DBpedia Person SHACL Shape},label={lst:Disshape}]
@prefix dbo: <http://dbpedia.org/ontology/> .
@prefix sh: <http://www.w3.org/ns/shacl#> .

ex:DBpediaPerson a sh:NodeShape;
 sh:targetClass foaf:Person;  
# node type Literal
 sh:property [sh:path foaf:name; 
  sh:minCount 1;        
  sh:nodeKind sh:Literal ];
# for MIN1 and MAX1 cardinality   
 sh:property [ sh:path dbo:birthDate;
  sh:datatype xsd:date ; 
  sh:minCount 1;   
  sh:maxCount 1;   
  sh:nodeKind sh:Literal ] ;
# node type IRI
 sh:property [sh:path dbp:birthPlace; 
  sh:nodeKind sh:IRI;
  sh:or ( [sh:class schema:Place] 
    [ sh:class dbo:Place ] )
  ];
# node type literal
 sh:property [ sh:path dbp:deathDate;
  sh:nodeKind sh:Literal;
  sh:datatype xsd:date ] .
  
\end{lstlisting}


\subsection{Limitations and Future Work} \label{sec:DisLimitation}

In this section, we discuss the limitations of the proposed approach, together with future research directions. 

\textit{Impact of addition of entities.} A limitation of the current approach is that we only considered the negative impact of deletion of entities as causes of quality issues. As a future research direction, we plan to study how to dynamically adapt impact of the addition of entities in an evolving KB. Furthermore, we argue that quality issues can be identified through monitoring lifespan of a KB. This argument has led to conceptualize the stability feature, which is meant to detect anomalies in a KB. Using a simple linear regression model, we explore the lifespan of an entity type. We can envision that stability feature can be used for analyzing the impact of the addition of entities. As a future work, we plan to monitor various KB growth rates to explore stability feature. In particular, we want to investigate further \textit{(i)} which factors are affecting stability feature, and \textit{(ii)} validating the stability measure.

\textit{Schema based validation.} We presented experimental analysis using three constraints types: cardinality, range, and string. As a future work, we plan to extend our implementations to other SHACL constraints. 
We envision that these constraints can be applied to other tools such as RDFUnit~\cite{kontokostas2014test} as a direct input. However, in RDFUnit they considered constraints in the form of RDFS/OWL axioms. We considered extending our approach to RDFUnit as future research work to favor the interoperability.

Furthermore, in our experimental analysis, we involved a human annotator to validate the datasets in order to create the partial gold standards. As future work, we plan to extend our evaluation strategy with an alternative approach such as the validation using OWL schema. However, it is challenging to explore an OWL schema for validation tasks. For example, the DBpedia KB 201610 version ontology lacks axioms about cardinality constraints (owl:cardinality, owl:minCardinality, maxCardinality). The only information that we can extract from the ontology is indirectly using the axioms that define functional properties (i.e., MAX1 constraints). In this context, we plan to extend our approach to other KBs that contain complete OWL schema representations.

\section{Conclusions}\label{sec:conclusion}


The primary motivations of this work are rooted in the concepts of Linked Data dynamics on the one side and constraints based KB validation on the other side. We focused on automatic shape validation as well as automating the timely process of quality issue detection without user intervention based on KB evolution analysis.  
Knowledge about Linked Data dynamics is essential for a broad range of applications such as effective caching, link maintenance, and versioning~\cite{kafer2013observing}. However, less focus has been given towards understanding knowledge base resource changes over time to detect anomalies over various releases. 
We introduced a completeness analysis approach by analyzing the evolution of a KB, to understand the impact of linked data dynamicity. More specifically, we explored the completeness of an entity type using periodic data profiling. However, we perceive that if the KB has design issues, our completeness analysis might lead to increase the number of false positives. We introduced an RDF validation approach to explore the consistency of KB resources using integrity constraints from SHACL representation. Our approach follows a traditional data mining workflow: data collection, data preparation, and model training. This approach can be applied to any knowledge base, and we demonstrated its usages for two different use cases, namely 3cixty Nice and DBpedia. We summarized the main findings of this work as follows:


In response to RQ1, the proposed approach provides an assessment of the overall completeness quality characteristic and 
it aims to identify potential problems in the data processing pipeline. Such an approach produces a smaller number of coarse-grained issue notifications that are directly manageable without any filtering and provide useful feedback to data curators. An experimental analysis of proposed completeness analysis is performed on two different knowledge bases of different size and semantics, and its operations are verified using these use cases. 
Since this approach uses simple statistical measures (count and difference), it reduces the search space of the suspicious issues, resulting in an approach that can be applied to also larger knowledge bases. 
Based on the two use cases, completeness analysis has proven to be highly effective to identify quality issues 
in the data extraction and integration process. Overall, the proposed approach achieved the precision of greater than 90\% for completeness measures for almost all use cases.
    
To address RQ2, a consistency analysis approach is proposed using constraints based feature extraction and learning models. This approach is evaluated using cardinality, and range constraints. 
In the experimental analysis, the performance of the five learning models are empirically assessed and the best performing model is identified according to the F1 score. The proposed approach reaches an F1 score greater than 90\% with DBpedia datasets for cardinality constraints using Random Forest model. Nevertheless, the proposed approach is defined in a generic and flexible manner which can be extended to other types of constraints. Overall, all learning models have good performances meaning that the problem is well configured and the features are predictive. 

\section*{Acknowledgements}
\noindent
This research has been partially supported by the \textit{4V: Volumen, Velocidad, Variedad y Validez en la gestin innovadora de datos (TIN2013-46238-C4-2-R)} project with the BES-2014-068449 FPI grant.

\bibliographystyle{model1-num-names}
\bibliography{sample}

\end{document}